\documentclass[%
reprint,
superscriptaddress,
amsmath,amssymb,
aps,
]{revtex4-1}

\usepackage{graphicx}
\usepackage{dcolumn}
\usepackage{bm}
\usepackage{mathtools}
\usepackage[caption=false]{subfig}

\begin{document}


\title{Origin and evolution of ferroelectricity in the layered rare-earth-titanate, $R_2$Ti$_2$O$_{7}$, Lichtenberg phases}

\author{Maribel N\'u\~nez Valdez}
\email{mari\_nv@gfz-potsdam.de}
\affiliation{%
 Materials Theory, ETH Z\"urich, Wolfgang-Pauli-Strasse 27, CH-8093 Z\"urich, Switzerland 
}%
\affiliation{Deutsches GeoForschungsZentrum GFZ, Telegrafenberg, 14473 Potsdam, Germany}
\affiliation{Goethe-Universit\"at Frankfurt, Altenh\"oferallee 1, 60438 Frankfurt am Main, Germany}
\author{Nicola A. Spaldin}%
\email{nicola.spaldin@mat.ethz.ch}
\affiliation{%
 Materials Theory, ETH Z\"urich, Wolfgang-Pauli-Strasse 27, CH-8093 Z\"urich, Switzerland 
}%

\date{\today}
\begin{abstract}
We report a systematic first-principles study based on density functional theory (DFT) of the structural and ferroelectric properties of the $R_2$Ti$_2$O$_{7}$ perovskite-related oxides with $R=$ La, Ce, Pr, and Nd. We show that, in all cases, the ferroelectric distortion to the ground-state polar $P2_1$ structure from its parent centrosymmetric $P2_1/m$ phase is driven by a single polar soft mode consisting of rotations and tilts of the TiO$_6$ octahedra combined with displacements of the $R$ ions. A secondary centrosymmetric distortion, which is stable in the parent structure, contributes substantially to the energy lowering of the ground state.  We evaluate the trends in structure and polarization across the series as a function of $R$ and reconcile discrepancies in reported values of polarization in the literature. Our results confirm that the family of $R_2$Ti$_2$O$_{7}$ materials belong to the class of proper geometric ferroelectrics. 
\end{abstract}

\maketitle
\section{Introduction and Background}
Perovskite-structure titanates, with chemical formula $A$TiO$_3$, are of tremendous fundamental and technological interest. In those with divalent $A$-site cations, in which the Ti ion is in the formally $d^0$ 4+ oxidation state, the properties range from quantum paraelectricity (in SrTiO$_3$) to large and robust ferroelectricity (in BaTiO$_3$ and PbTiO$_3$) and even to multiferroicity (in strained EuTiO$_3$). The $d^1$ configuration on the formally Ti$^{3+}$ ions in the rare-earth titanates such as LaTiO$_3$ leads to strong correlations and the associated Mott physics including magnetism and metal-insulator transitions.

\begin{figure}
\centering
    \subfloat[]{
        \includegraphics[width=0.47\textwidth]{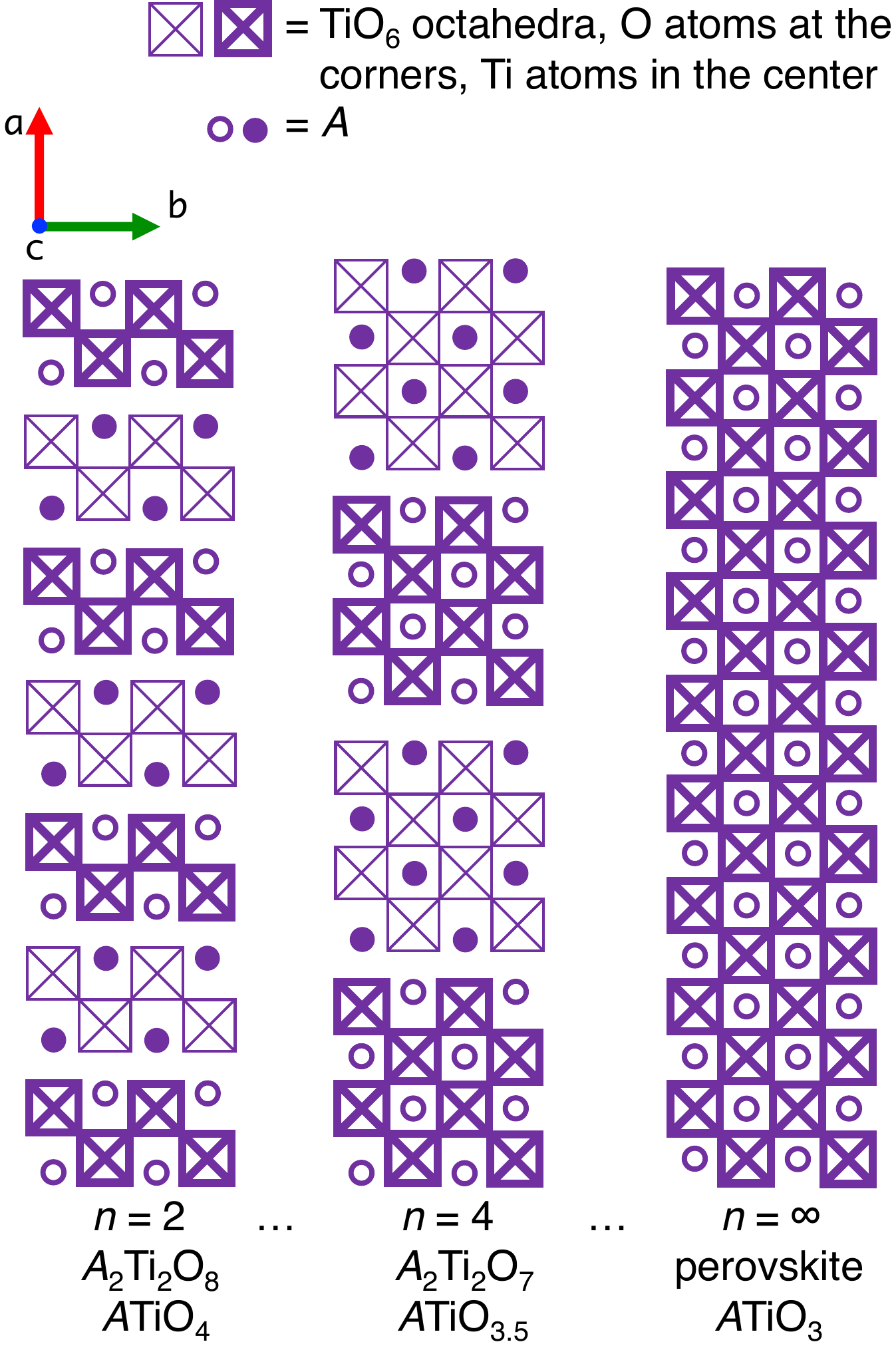}
        \label{fig1a}
        }\\
    \subfloat[]{
        \includegraphics[width=0.47\textwidth]{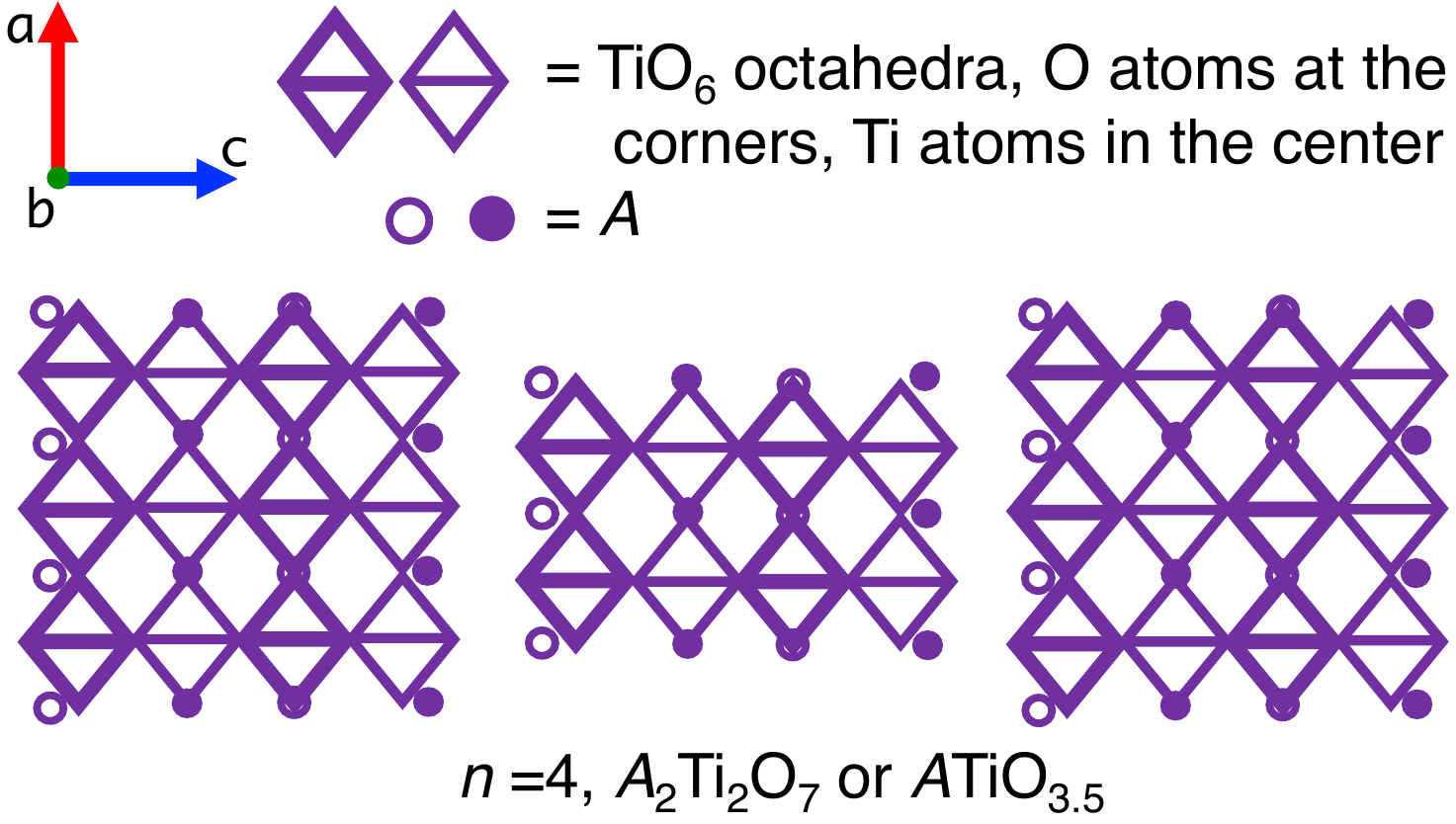}
        \label{fig1b}
        }
     \caption{(a) Crystal structures of the $n=2, 4$ and $\infty$ members of the layered perovskite-related $A_n$Ti$_n$O$_{3n+2}$ Lichtenberg phases viewed along the $a$ axis. Thick TiO$_6$ octahedra  and filled circles have a $\frac{1}{4}$ unit cell (half a simple perovksite cube) height difference along $a$ from the thin octahedra and unfilled circles. (b) Crystal structure of the $n=4$ material along the $b$ axis showing the 1D-chain-like corner-shared TiO$_6$ octahedra along the $a$ axis.}\label{fig1}
\end{figure}

The $A_n$Ti$_n$O$_{3n+2}$ family of materials, or {\it Lichtenberg phases} \cite{frank}, of which the perovskite structure is the $n=\infty$ end member, offers additional structural flexibility and in turn additional functionality within this crystal chemistry class (for a thorough review see Ref.~[\onlinecite{frank}] and references therein). They are obtained by cutting the cubic perovskite $A$TiO$_3$ structure perpendicular to the [110] direction into slices that are $n$ perovskite unit cells thick, and then inserting additional compensating oxygen ions to complete the oxygen octahedra on either side of the cut (110) planes. The structures obtained for $n=2, 4$ and $\infty$  are shown in Fig. \ref{fig1}(a). Many different properties have been reported within this family, depending on the identity of the $A$-site cation and the value of $n$, for example quasi-one-dimensional metallicity in the $n=5$ LaTiO$_{3.4}$, and structural complexity in well-ordered multi-layered stacking sequences in $n=4\frac{1}{3}$ Ca-doped LaTiO$_{3.46}$ \cite{1981Nanot}.  
%
Of particular interest are the $n=4$ members with trivalent rare-earth $A$ sites, $R_4$Ti$_4$O$_{14}$, usually written as $R_2$Ti$_2$O$_{7}$ or $R$TiO$_{3.5}$. For this chemistry, The $n=4$ layered perovskite phase is the stable phase at ambient pressure for ratios of $R^{3+}$ to Ti$^{4+}$ cation radii of $r_{R^{3+}}/r_{\text{Ti}^{4+}}\ge 1.80$ \cite{saha}, and therefore occurs for $R=$ La, Ce, Pr, and Nd; smaller rare earths form the cubic pyrochlore structure \cite{pyrochlore}. In addition, $R_2$Ti$_2$O$_{7}$ layered perovskites with $R$= Sm, Eu, and Gd can be stabilized using high pressure synthesis \cite{1987Titov}. (Under thin-film growth conditions, a metastable (001)-oriented layered structure has been reported for Sm$_2$Ti$_2$O$_7$, Eu$_2$Ti$_2$O$_7$ and Gd$_2$Ti$_2$O$_7$ \cite{2012Shao,2012Shaob}.) The $n=4$  materials are ferroelectric with Curie temperatures, $T_C$, in the range 1600-1850 K, among the highest known. As a result, they are of interest for high-temperature applications such as piezoelectric transducers \cite{damjanovic}, acoustic and vibration sensors \cite{turner}, high-$k$ dielectrics \cite{atuchin}, or as photocatalysts \cite{hwang}.  The goal of this work is to use first-principles electronic structure calculations based on density functional theory to clarify the origin of the ferroelectricity and to extract and rationalize the trends in the structural and functional properties across the La$_2$Ti$_2$O$_{7}$ - Ce$_2$Ti$_2$O$_{7}$ - Pr$_2$Ti$_2$O$_{7}$ - Nd$_2$Ti$_2$O$_{7}$ series.
\subsection{Crystal Symmetry of $A_2$Ti$_2$O$_{7}$}
The $R_2$Ti$_2$O$_{7}$ Lichtenberg phases with $R=$ La, Ce, Pr, Nd are monoclinic with the polar $P2_1$ space group at ambient conditions. The earliest reports on growth and crystal symmetry are for La$_2$Ti$_2$O$_7$ (LaTO)~\cite{nana} and Nd$_2$Ti$_2$O$_7$ (NdTO)~\cite{kimura}. LaTO and NdTO crystals of $\sim$5~mm in diameter and $\sim$50~mm in length were grown using the floating zone technique and analyzed using X-ray diffraction (XRD). These LaTO and NdTO crystals were shown to belong to the monoclinic system $P2_1$ with four formula units (f.u.) per unit cell (u.c.). More recently, CeTi$_2$O$_{7}$ (CeTO) (using powder XRD) and Pr$_2$Ti$_2$O$_{7}$ (PrTO) (from XRD analysis of pellets) were identified to be isostructural with LaTO and NdTO \cite{2008Kim,2013Sun}. The measured lattice parameters and monoclinic angles for all four materials are given in Table~\ref{table1}.

Additionally, it has been reported that LaTO undergoes a high-temperature transformation at $\sim$1053~K from its room-temperature monoclinic phase to a polar orthorhombic phase of symmetry $Cmc2_1$ also with four formular units per unit cell~\cite{ishizawa2}. The lattice parameters at 1173 K, measured using XRD were found to be $a=3.95$~\AA, $b=5.61$~\AA, and $c=25.95$~\AA~, that is approximately halved along the $a$ axis and doubled along $c$ compared with the $P2_1$ phase. Note that this $Cmc2_1$ phase is polar, and so can not be the paraelectric parent of the ferroelectric structure. It has not been observed for CeTO, PrTO or NdTO.
\subsection{Ferroelectricity of $A_2$Ti$_2$O$_{7}$}
There have been some experimental and computational efforts towards measuring and understanding the ferroelectric properties of the titanate Lichtenberg phases, in particular for the cases of LaTO, NdTO, and to a lesser degree for PrTO and CeTO. From dielectric measurements, single crystals of monoclinic $P2_1$ LaTO have been reported to be ferroelectric with spontaneous polarizations, $P_S$, of 5~$\mu$C/cm$^2$ and 9~$\mu$C/cm$^2$ \cite{nana,kimura}. For PrTO nanocrystalline pellets, a $P_S$ of $\sim$7~$\mu$C/cm$^2$ was observed \cite{2013Sun}, while a polycrystalline sample of PrTO  yielded $P_S=0.017$~$\mu$C/cm$^2$ \cite{2015Patwe}, however both works acknowledged a lack of saturation of the polarization-electric field loops due to leakage currents. Values of $P_S=4.4$, 1.8 and 4.19~$\mu$C/cm$^2$ for thin films of LaTO, NdTO \cite{2002Kim} and CeTO \cite{2008Kim} respectively, have been reported. In other studies addressing the ferroelectric properties of LaTO, NdTO, PrTO and CeTO films \cite{2013Bayart,2014Bayart,2016Bayart}, the authors declared the existence of ferroelectricity from observed piezoloops, but no values of $P_S$ were given. 
\begin{figure}
\includegraphics[width=0.35\textwidth]{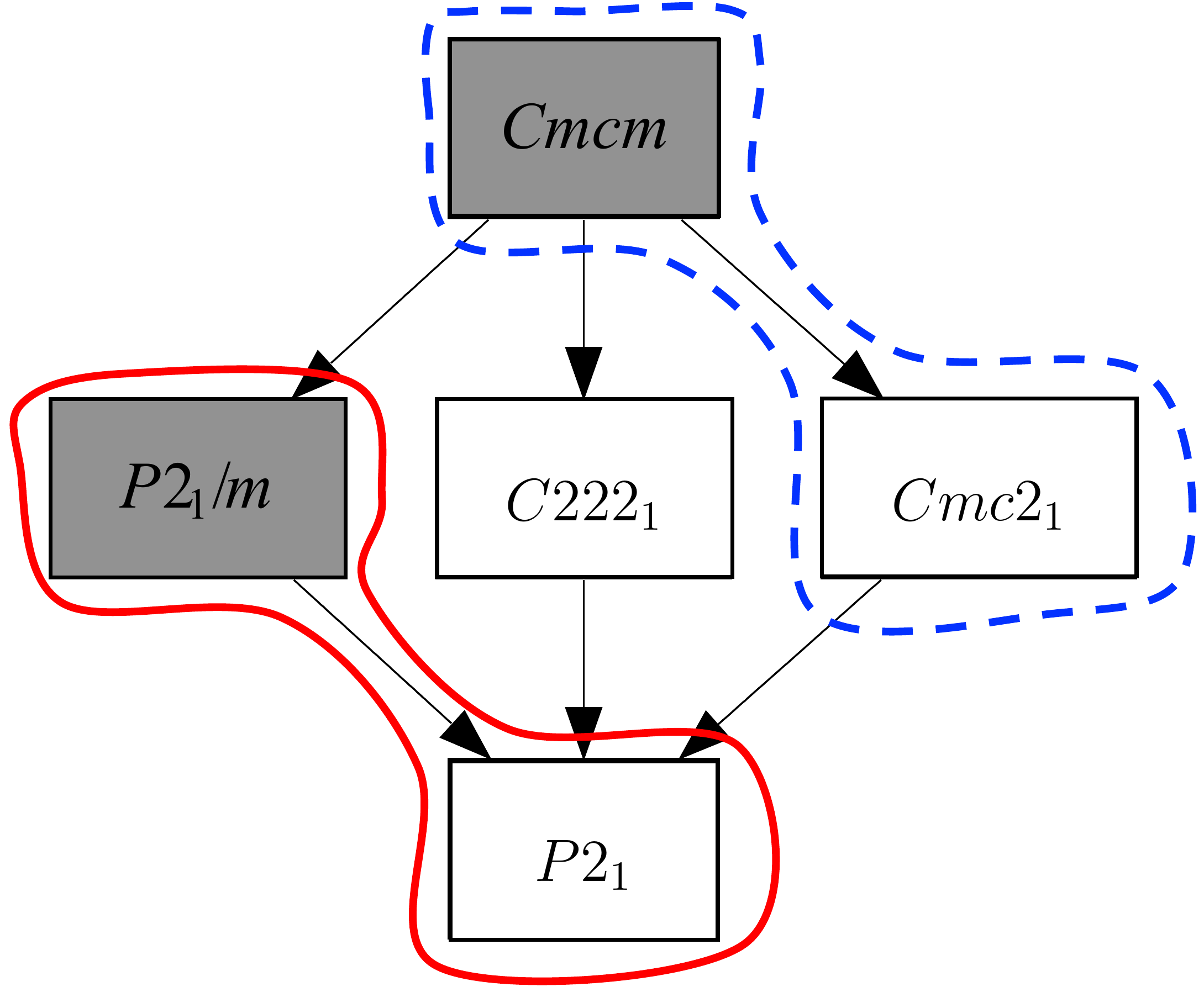}
\caption{\label{fig02}Group-subgroup decomposition showing the possible symmetries in the ferroelectric-paraelectric transformation of the perovskite-related layered $R_2$Ti$_2$O$_{7}$ materials. Centrosymmetric space groups (non-polar structures) are shaded gray, polar space groups are white. The pathway studied in this work (enclosed by the red solid line), as well as that in Ref.~\cite{lopez} (enclosed by the dashed blue line) are indicated. 
}
\end{figure}

Regarding the nature of the transition to the ferroelectric phase, in Fig.~\ref{fig02} we show a group-subgroup analysis \cite{2000Ivantchev} using the Bilbao Crystallographic Server \cite{2006aAroyo,2006bAroyo,2011Aroyo} indicating possible pathways for reaching the reported $P2_1$ ground-state structure. The analysis indicates one direct single-step possibility for reaching the $P2_1$ ground state from a non-polar structure, which is via the $P2_1/m\rightarrow P2_1$ transition. 
While the $P2_1/m$ centrosymmetric phase has not been reported experimentally for the titanate Lichtenberg phases, it has been observed for the $n=4$ tantalate Sr$_2$Ta$_2$O$_{7}$ \cite{2004Hushur}. 
It is this transition that we explore in this work. We also see that a direct transition from the polar $Cmc2_1 $ phase mentioned above to the $P2_1$  ground state is indeed allowed, and that this pathway would imply a $Cmcm$ paraelectric phase. 
An earlier first-principles theoretical study addressed the nature of the non-polar to polar $Cmcm \rightarrow Cmc2_1$ transition \cite{lopez} and found that the leading instability causing the transition consists primarily of TiO$_6$  octahedral rotations, which occur because the $A$-site cation is too small to maintain the high-symmetry structure. The layered connectivity of the crystal structure means that, in contrast to the case of the three-dimensionally connected perovskites, such rotational modes can be polar. The mechanism (called {\it topological} ferroelectricity by the authors of Ref.~[\onlinecite{lopez}]) is analogous to the {\it geometric} ferroelectric mechanism previously identified in the Ba$M$F$_4$ family ($M$=Mn, Fe, Co, or Ni) \cite{ederer}, which have the same structure as the $n=2$ member of the $A_nB_n$O$_{3n+2}$ series. It is strikingly different from the conventional ferroelectricity in related perovskite-structure oxides such as BaTiO$_3$, which occurs through off-centering of the Ti ion from the center of its oxygen octahedron, and is driven by the resulting enhanced covalent-bond formation with the closer oxygen ion(s) \cite{ederer,lopez,2016MNV}. 

\begin{figure}
\includegraphics[width=0.5\textwidth]{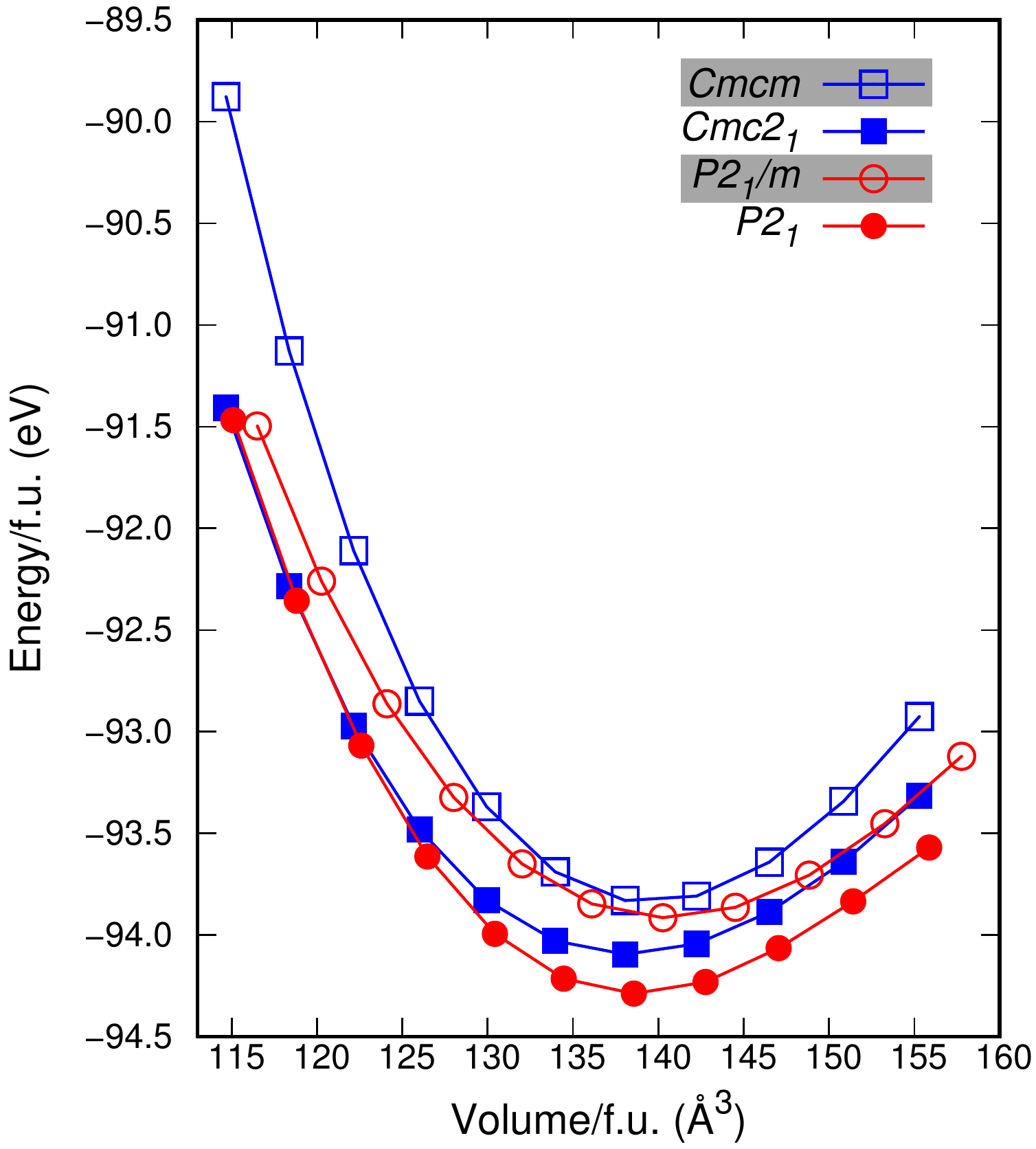}
\caption{\label{fig03} Total energy per f.u. as a function of volume per f.u. for Ce$_2$Ti$_2$O$_7$ for the two polar (filled symbols) and two non-polar (empty symbols) phases discussed in this work. }
\end{figure}

Total energy calculations have confirmed the experimental observation that the $P2_1$ phase is the most stable structure for LaTO, NdTO \cite{bruyer} and PrTO \cite{2015Patwe}, and found the centrosymmetric $P2_1/m$ phase to be lower in energy than the centrosymmetric $Cmcm$ structure. 
In Fig.~\ref{fig03} we show our calculated total energies as a function of volume for the relevant phases of CeTO, where we again find that the lowest energy polar and non-polar phases are $P2_1$ and $P2_1/m$ respectively, further motivating our investigation of the details of this transition in this work.

A number of first-principles calculations of the magnitude of the spontaneous polarization for the observed $P2_1$ phase, taking the non-polar $P2_1/m$ phase as reference structure,have been performed, and $P_S$ magnitudes of  7.72, 7.42, and 8.3~$\mu$C/cm$^2$ for LaTO, NdTO \cite{bruyer}, and PrTO \cite{2015Patwe}, respectively, were obtained.   
Interestingly, the spontaneous polarization calculated in Ref.~[\onlinecite{lopez}] for the $Cmcm \rightarrow Cmc2_1$ in LaTO was $P_{S}=29~\mu$C/cm$^2$, almost four times larger than the previously reported calculated values for the $P2_1/m \rightarrow P2_1$ transition \cite{bruyer}. It was argued, however, that a direct comparison between the $P_S$ values for  the $P2_1$ and $Cmc2_1$ phases might not be straightforward,  and that although not typical of FE perovskites, the further transformation $Cmc2_1\rightarrow P2_1$ could involve a reduction in $P_S$ of LaTO \cite{lopez}.

In this work we provide a systematic investigation of the rare-earth titanates $R_2$Ti$_2$O$_{7}$ for $R=$ La, Ce, Pr and Nd using density functional theory (DFT). Our goals are first to understand the ferroelectric mechanism with a particular focus on explaining the discrepancies in the reported values of polarization, and second to facilitate a systematic comparison between the members of the $R_2$Ti$_2$O$_7$ family by revealing trends across the series. We single out Ce$_2$Ti$_2$O$_{7}$ for detailed analysis, as its calculated ferroelectric properties have not been reported previously. 

The remainder of this paper is organized as follows. First, we briefly describe the computational methods we used in our calculations. Then we present and discuss our results for the structural, electronic and ferroelectric properties of the $R_2$Ti$_2$O$_7$ systems for $R=$ La, Ce, Pr, and Nd. In particular, we focus on the ferroelectric mechanism underlying the polar structural distortions. In the final section we summarize our conclusions. 
\begin{figure}
\includegraphics[width=0.49\textwidth]{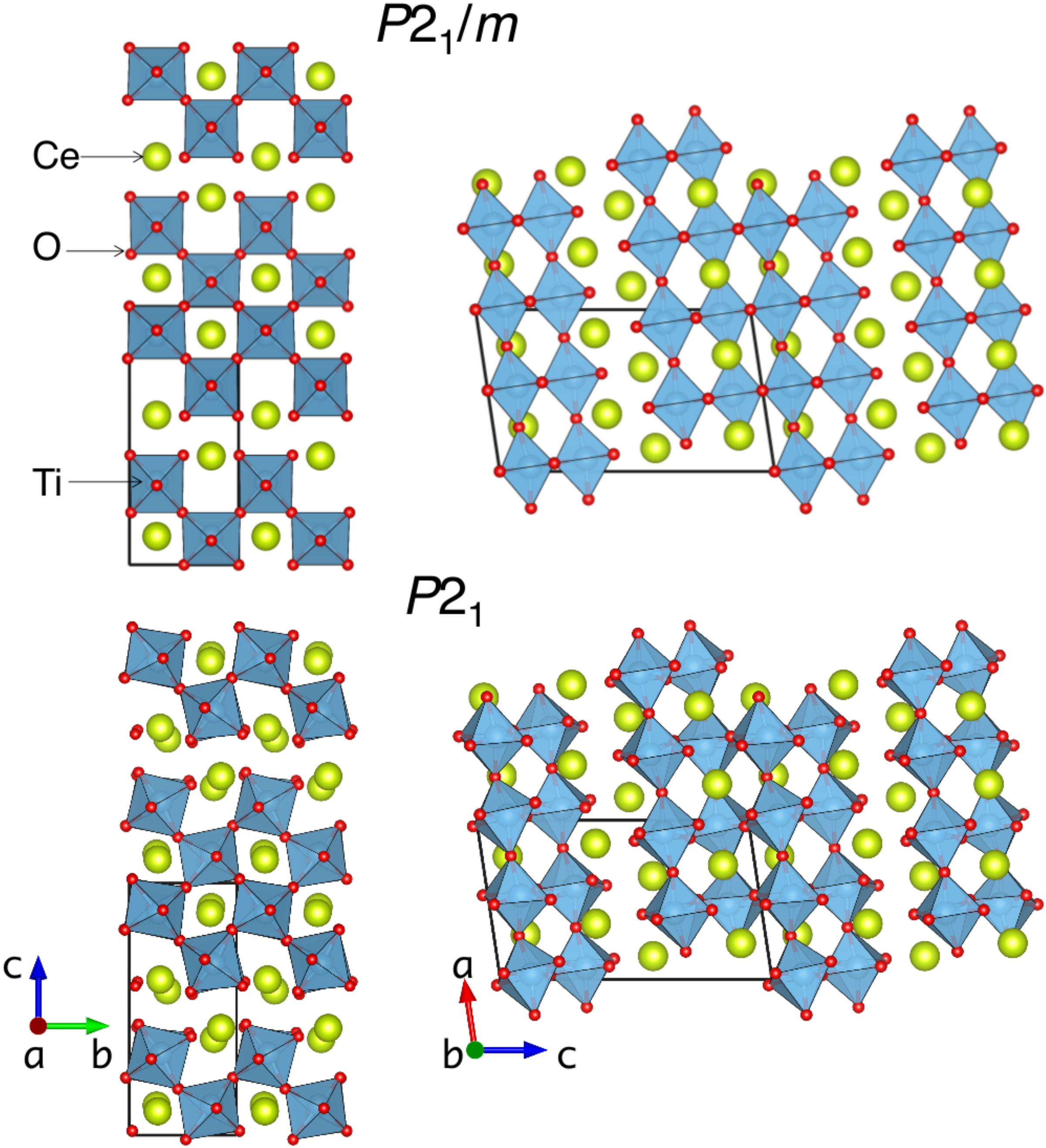}%
\caption{\label{fig04}Calculated high-symmetry centrosymmetric $P2_1/m$ and low-symmetry polar $P2_1$ structures of Ce$_2$Ti$_2$O$_{7}$. The 44-atom unit cells (four formula units) are indicated by the black lines. The calculated structures of $R_2$Ti$_2$O$_{7}$ with $R=$ La, Pr, and Nd are similar. 
}
\end{figure}
\section{Computational Details}
Our first-principles calculations are based on density functional theory (DFT) \cite{dft1,dft2} using the Vienna \textit{Ab initio} Simulation Package (VASP) \cite{kresse93,kresse94,kresse96a,kresse96b}. We use the generalized gradient approximation (GGA) \cite{perdew1} in the Perdew, Burke and Ernzerhof (PBE) \cite{perdew2} format revised for solids (PBEsol) \cite{PBEsol} for the exchange-correlation functional as it gives better agreement between the optimized and experimental lattice parameters for the systems under consideration compared with the standard PBE functional. We use the default PAW potentials \cite{blochl,kresse99}, including six valence electrons for the oxygen (2s$^2$2p$^4$), four for Ti (3d$^3$4s$^1$) and 11 for the $R$ atoms (5s$^2$5p$^6$5d$^1$6s$^2$), with the $f$ electrons treated as core states. The cutoff energy for the plane-wave expansion of the wave-functions is 550 eV and 4$\times$4$\times$2 and 7$\times$1$\times$5 Monkhorst-Pack \cite{kmesh} $\mathbf{k}$-point meshes are used for the Brillouin zone sampling of the monoclinic and orthorhombic phases, respectively. In the structural optimizations, we fully relax the 44-atom unit cells, lattice parameters and ions, until all forces are converged to less than 0.1 meV/\AA\ on each atom. For the calculation of the spontaneous polarization ($P_S$) we use the Berry-phase formalism \cite{berry1,berry2,berry3,spaldin} with integration along ten homogeneously distributed strings each of ten $\mathbf{k}$-points parallel to the reciprocal $b$ axis in the Brillouin zone. Our reported $P_S$ values are the differences in polarization between the high-symmetry centrosymmetric reference structures belonging to the $P2_1/m$ space group and the corresponding low-symmetry ferroelectric (FE) $P2_1$ structures along the same branch in the polarization lattice.
\begin{figure}
\includegraphics[width=0.5\textwidth]{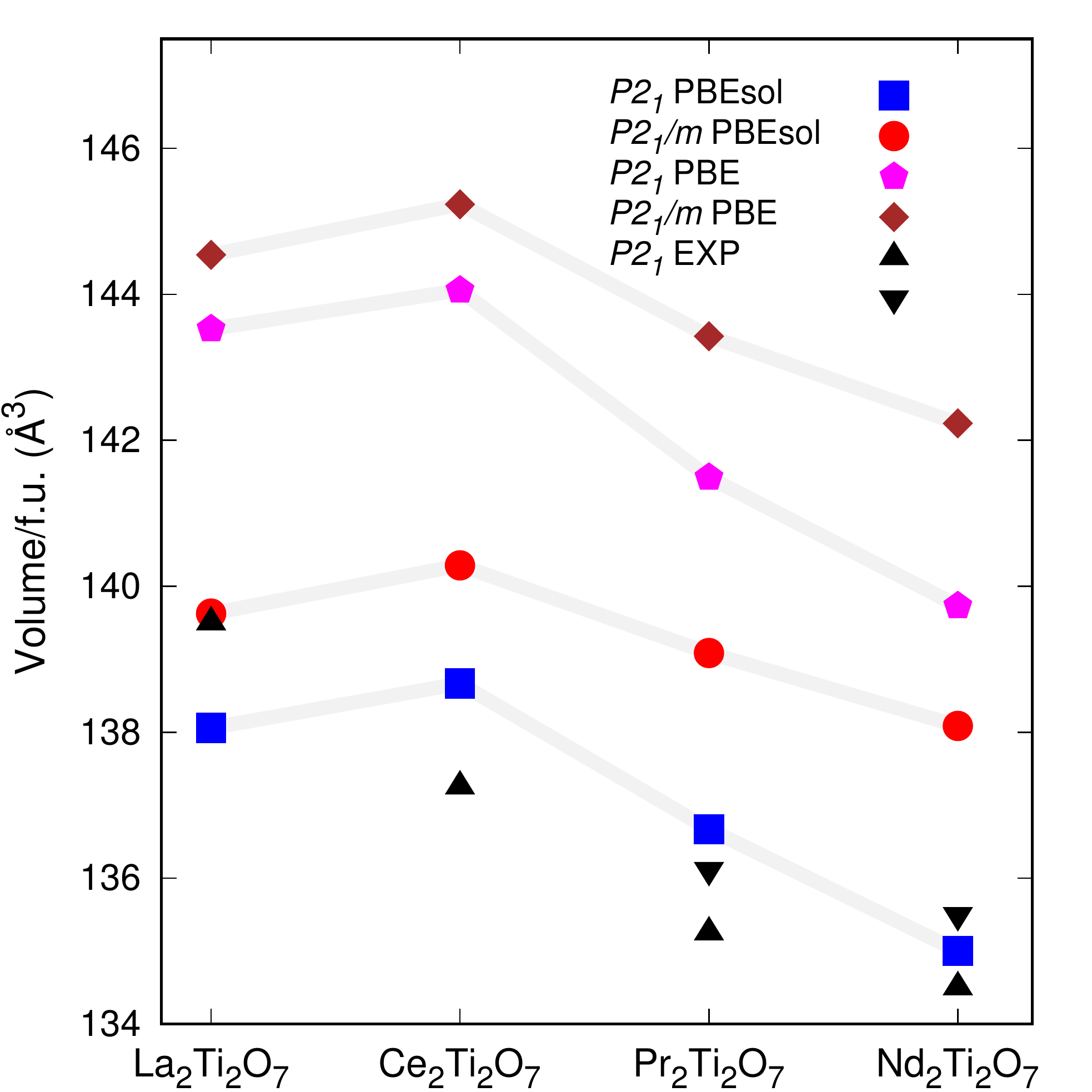}
\caption{\label{fig5} Volume per f.u. of the titanate Lichtenberg phases studied in this work calculated using the PBEsol and PBE functionals as well as measured experimentally (EXP: LaTO \cite{nana}, CeTO \cite{frank}, PrTO \cite{2015Patwe,frank}, and NdTO~\cite{frank,kimura}). We see that the PBEsol functional provides the best match to the experimental volumes.}
\end{figure}
\begin{table*}
\caption{\label{table1} Lattice parameters ($a$, $b$, and $c$), monoclinic angle ($\beta$), and volume ($V$) of polar ($P2_1$) and centrosymmetric ($P2_1/m$) phases of $R_2$Ti$_2$O$_7$ compounds for $R=$ La, Ce, Pr, and Nd computed using the PBEsol functional compared to the standard PBE and experiment. Results from this work are labeled by [*] and experimental  parameters by EXP. PW stands for the Perdew-Wang parametrization.}
\begin{ruledtabular}
\begin{tabular}{lllllcccccccc}
\multicolumn{3}{r}{} &  \multicolumn{2}{c} { $a$ (\AA) } & \multicolumn{2}{c} {$b$ (\AA)} & \multicolumn{2}{c} {$c$ (\AA)} & \multicolumn{2}{c} {$\beta$ (deg)} & \multicolumn{2}{c} {$V$ (\AA$^3$)} \\
\cline{3-13}
\multicolumn{2}{r}{}& Ref. & $P2_1$ & $P2_1/m$ & $P2_1$ & $P2_1/m$ & $P2_1$ & $P2_1/m$ & $P2_1$ & $P2_1/m$ & $P2_1$ & $P2_1/m$ \\
\colrule
 La$_2$Ti$_2$O$_7$ & PBEsol & [*] & 7.77 & 7.82 & 5.51 & 5.50 & 13.05 & 13.14 & 98.58 & 98.55 & 552.23 & 558.50 \\
 
  & PBE& [*] & 7.81 & 7.90 &  5.60 &  5.56 &  13.28 &  13.32 & 98.50 & 98.49  & 574.10 & 578.16  \\ \\
 
  & PW & \cite{bruyer} & 7.82 & 7.89 & 5.59 & 5.56 & 13.26 & 13.31 & 98.51 & 98.51 & 573.16 & 577.55 \\
 
 & EXP & \cite{nana} & 7.81 &  &  5.55 &  & 13.02 &  & 98.70 & & 558.00 & \\
 
\colrule
Ce$_2$Ti$_2$O$_7$ & PBEsol & [*] &  7.75 & 7.83 & 5.52 & 5.50 & 13.10 & 13.16 & 98.56 & 98.55 & 554.66 & 561.14\\
 
                                    & PBE & [*] & 7.80  & 7.92  & 5.60 & 5.56 & 13.33 & 13.34 & 98.49 & 98.48 & 576.22 & 580.93 \\ \\
 
  & EXP & \cite{frank} & 7.76 & & 5.51 & & 12.99 & & 98.50 & & 549.00 & \\
\colrule
Pr$_2$Ti$_2$O$_7$ & PBEsol & [*] & 7.70 & 7.78 & 5.48 & 5.49 & 13.08 & 13.17 & 98.53 & 98.49 & 546.68 & 556.34 \\

                          & PBE & [*] & 7.74 & 7.86 & 5.56 & 5.54 & 13.30 & 13.33 & 98.47 & 98.56 & 565.96 & 573.70 \\ \\

 & PBE & \cite{2015Patwe} & 7.75 & 7.87 & 5.56 & 5.55 & 13.30 & 13.35 \\

 & EXP & \cite{frank} & 7.69 & & 5.47 & & 12.99 & & 98.40 & & 541.00 & \\

  & EXP & \cite{2015Patwe} & 7.72 & 7.71  & 5.49 & 5.49 & 13.00 & 13.00 & 98.55 & 98.53 & 544.47 & 544.40 \\

\colrule
Nd$_2$Ti$_2$O$_7$ & PBEsol & [*] & 7.66 & 7.74 & 5.45 & 5.47 & 13.06 & 13.18 & 98.51 & 98.45 & 540.01 & 552.34\\
                          & PBE & [*] & 7.70 & 7.81 & 5.52 & 5.52 & 13.29 & 13.33 & 98.43 & 98.56 & 558.92 & 568.93 \\ \\
                          
  & PW & \cite{bruyer} & 7.68 &  7.78  & 5.47 &  5.54  & 13.06 &  13.08 & 98.54 &   & 542.65 & 557.70 \\                          
 & EXP & \cite{frank} & 7.67 & & 5.46 & & 12.99 & & 98.50 & & 538.00 &\\
  & EXP & \cite{kimura} & 7.68 & & 5.48 & & 13.02 & & 98.28 & & 542.25 & 
\end{tabular}
\end{ruledtabular}
\end{table*}
\begin{figure}
\includegraphics[width=0.5\textwidth]{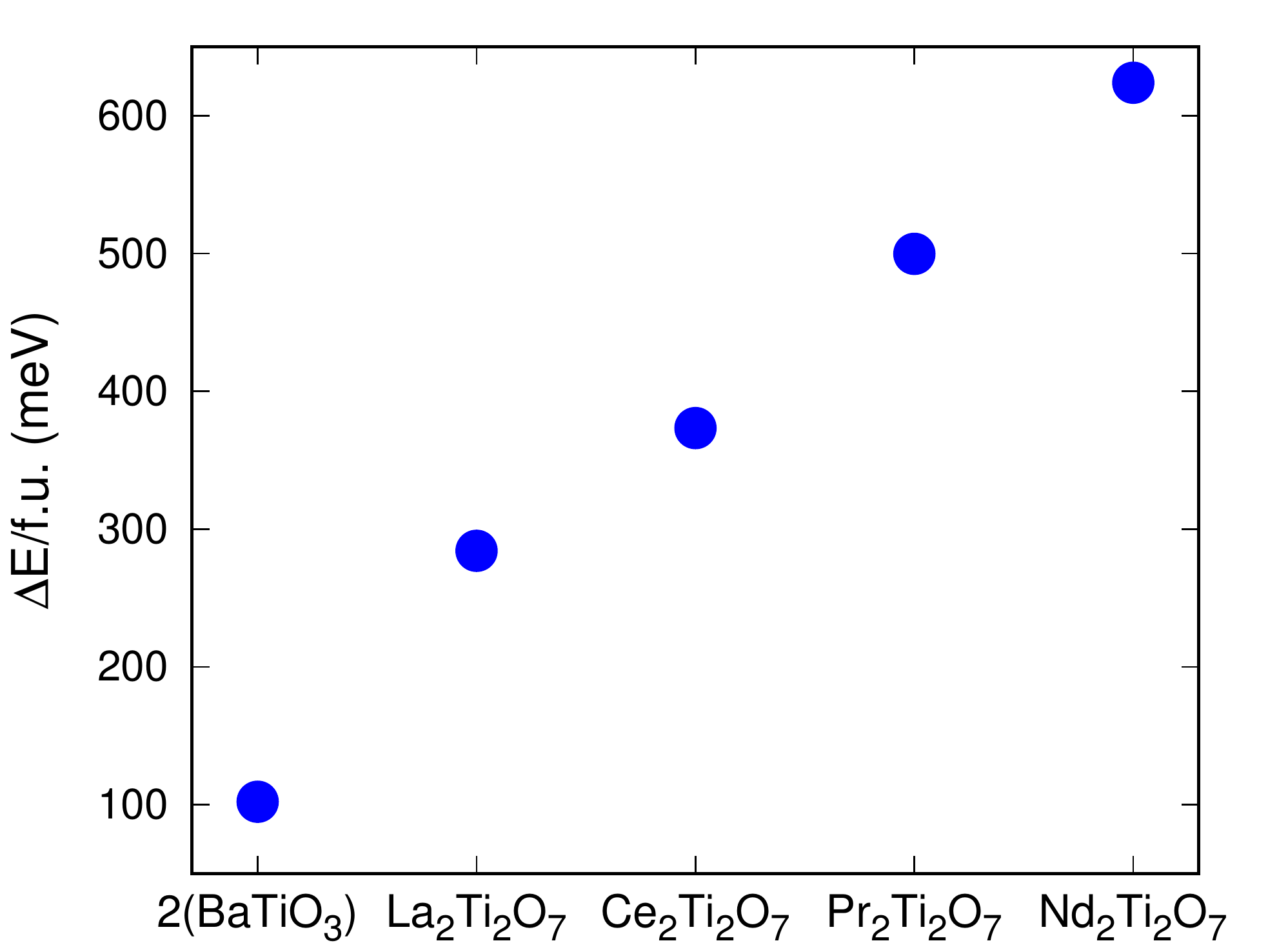}
\caption{\label{fig06} Difference in energy, $\Delta E$, per f.u. between the centrosymmetric $P2_1/m$ and polar $P2_1$ phases of the $R_2$Ti$_2$O$_{7}$ materials and the conventional ferroelectric BaTiO$_3$ (for consistency of comparison, the value for two formula units of barium titanate, 2(BaTiO$_3$), is shown) \cite{2017Zhang}.}
\end{figure}
\section{Results and Discussion}
\subsection{Energetics and Structural Properties} 
We begin by determining the lowest energy polar and non-polar structures for our series of rare-earth titanates. In Fig. \ref{fig03} we present the calculated total energy per formula unit as a function of volume for CeTO for the symmetries of the outer branches in the group-subgroup chart of Fig.~\ref{fig02}. We see clearly that the lowest energy polar phase has the monoclinic $P2_1$ space group, with the $Cmc2_1$ structure $\sim$193.0 meV/f.u. higher in energy, and that the lowest energy non-polar phase has $P2_1/m$ symmetry, $\sim$84.5 meV/f.u. lower in energy than the $Cmcm$ structure. The structures of these two lowest-energy phases, which consist of four formula units (44 atoms) per unit cell, are shown in Fig.~\ref{fig04}. Ce, Ti and O occupy 2$a$ Wyckoff positions in the $P2_1$ structure, while in the $P2_1/m$ centrosymmetric phase, Ce and Ti occupy 2$e$ positions and O splits into 2$a$, 2$e$ and 4$f$ Wyckoff positions.
We obtain the same energy ordering for the other members of the series, consistent with Refs.~[\onlinecite{2015Patwe}] and [\onlinecite{bruyer}]. 
 The $C222_1$ structure shown in Fig.~\ref{fig02}, has never been observed/reported for the titanate Lichtenberg phases or other $n=4$ chemistries, therefore it is not investigated here.

Our calculated lattice parameters, with both PBE and PBEsol functionals and for the $P2_1$ and $P2_1/m$ symmetries are listed in Table \ref{table1} for all systems and compared with available experimental data. In general, the agreement between our PBEsol results and measurements  is very good for $R=$ La, Ce, Pr, and Nd in $P2_1$ symmetry, with volume deviations of $\lesssim$1\% (for PBE, the deviations are up to $\sim$4.6\%). Fig. \ref{fig5} shows the volume per formula unit trends as a function of $R$. We notice that the experimental volume decreases in correspondence with the  contraction of the $R^{3+}$ ionic radius from La$\rightarrow$Ce$\rightarrow$Pr$\rightarrow$Nd. While this trend is captured by both the PBEsol and PBE functionals for the smaller rare earths, they both give anomalous results for the La compound, a behavior that has also been reported in calculations for the analogous $R_2$O$_3$ sesquioxides \cite{2003Hirosaki, 2007Wu}.  We also observe that, independently of functional, the volume decreases in the transformation from the non-polar $P2_1/m$ phase to the polar $P2_1$ phase. This volume reduction for the PBEsol optimized structures ranges between $\sim$1\% for La and $\sim$2\% for Nd. Given the better structural results rendered by PBEsol, we use this functional for the rest of this work.

A comparison of the energy  difference per formula unit, $\Delta E$/f.u., between the centrosymmetric $P2_1/m$ and ferroelectric $P2_1$ phases of $R_2$Ti$_2$O$_7$ for $R=$ La, Ce, Pr, and Nd is shown in Fig. \ref{fig06}.  This energy difference increases as the ionic radius of the $R$ cation and volume decrease. Also one can observe that the overall change in energy per formula unit associated with the ferroelectric $P2_1/m\rightarrow P2_1$ transition in $R_2$Ti$_2$O$_7$ compounds with $T_C$ between 1600 and 1850 K is considerably larger when compared, for example, to the corresponding energy change in two formula units of the conventional ferroelectric BaTiO$_3$, ($Pm\vec{3}m\rightarrow P4mm$, $T_C\sim400$ K)  \cite{2017Zhang}, in spite of BaTiO$_3$'s larger ferroelectric polarization. 
\subsection{Ferroelectric Properties}
\begin{figure}
    \centering
   \subfloat[]{
        \includegraphics[width=0.45\textwidth]{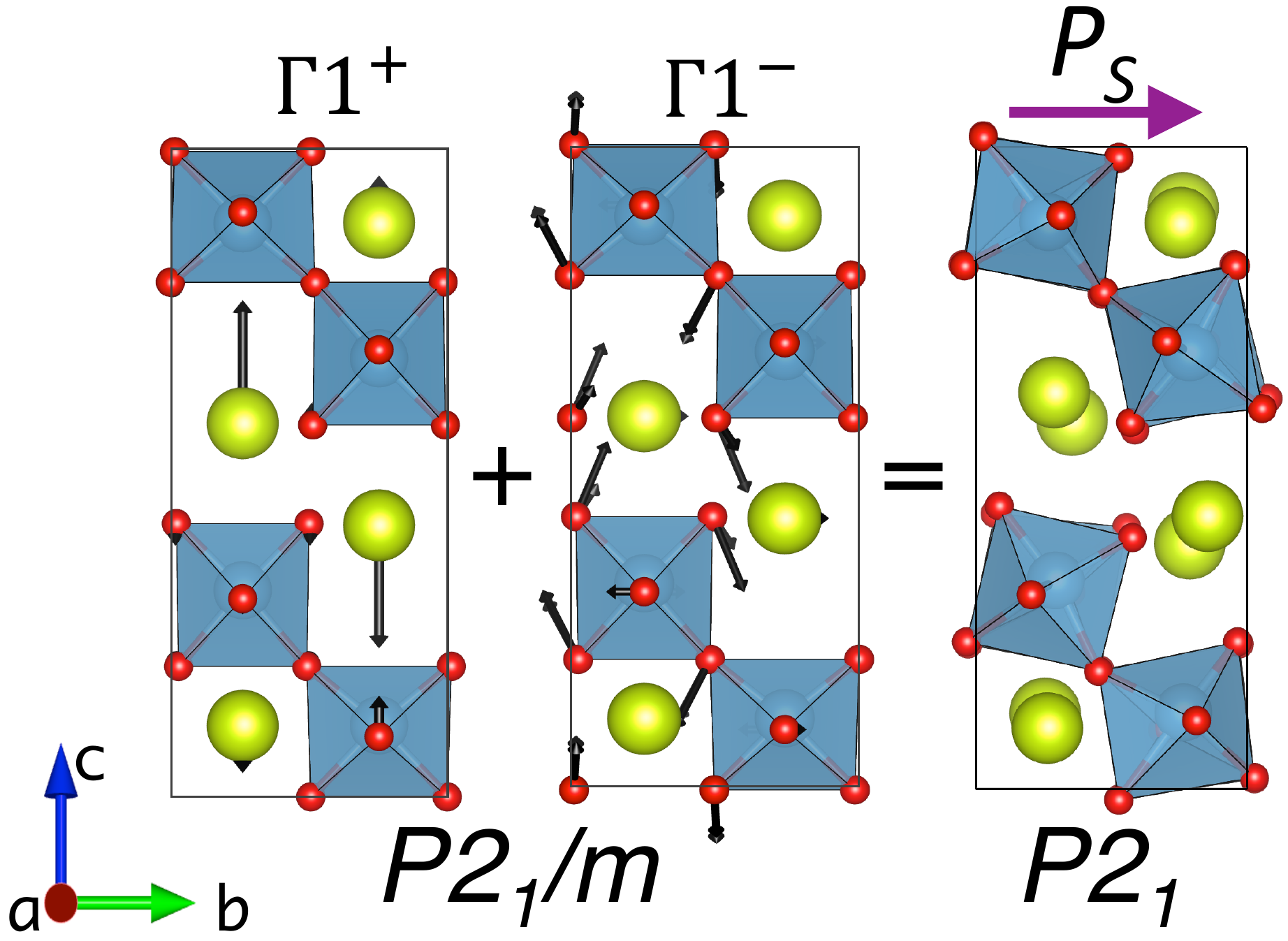}
        \label{fig7a}
    }\\
    \subfloat[]{
        \includegraphics[width=0.5\textwidth]{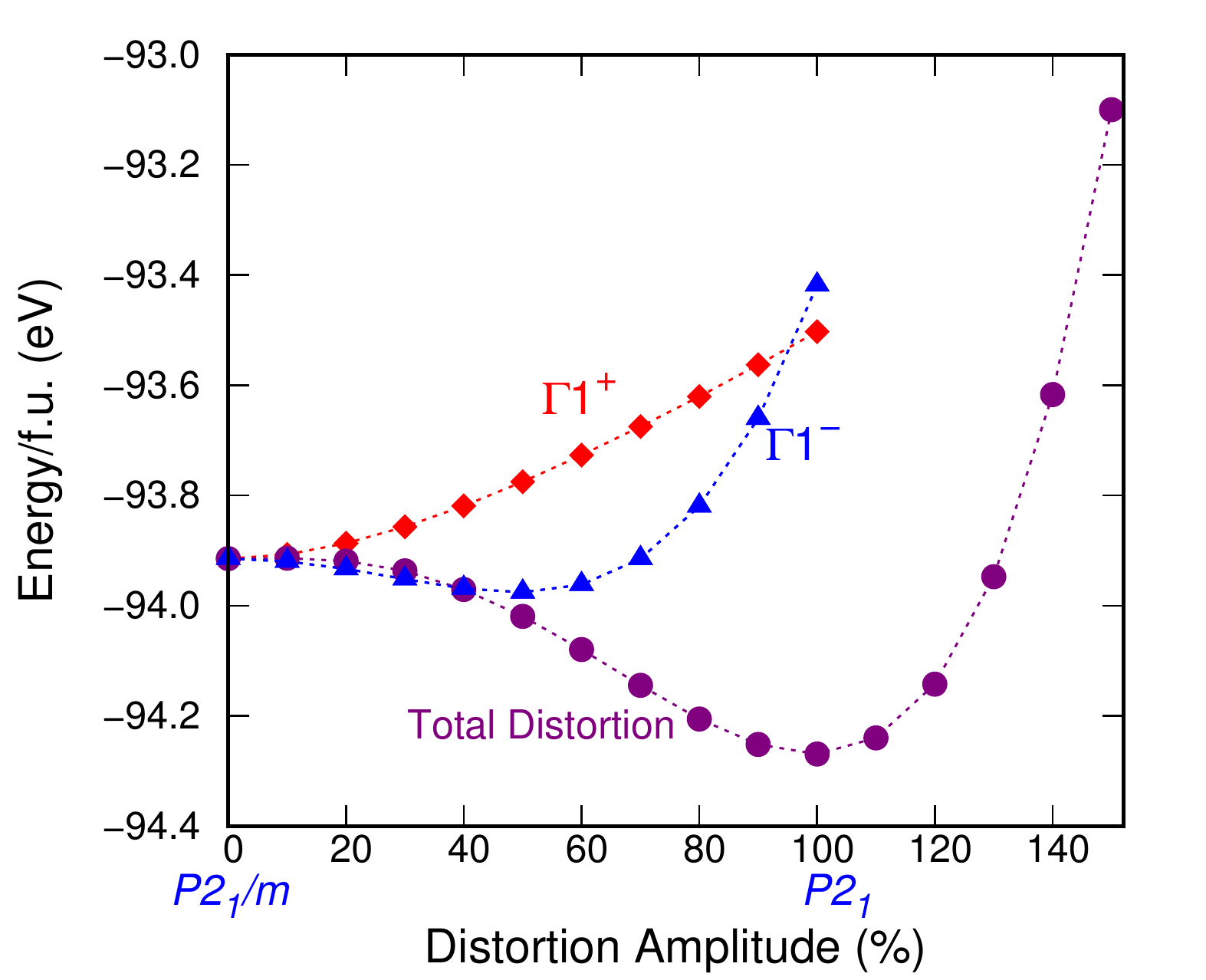}
        \label{fig7b}
    }
     \caption{(a) Representation of the mode decomposition in the $P2_1/m\rightarrow P2_1$ transformation of Ce$_2$Ti$_2$O$_7$. The arrows indicate the magnitude (doubled in length for visualization purposes) and direction of the atomic displacements, starting from the high-symmetry structure, in each mode. (b) Energy variation in Ce$ _2$Ti$_2$O$_7$ as a function of distortion from its high-symmetry $P2_1/m$ structure.  The purple points and line show the full structural distortion to the $P2_1$ ground state, whereas the red and blue lines indicate the change in energy when only the $\Gamma 1^+$ or the $\Gamma 1^-$ symmetry component of the distortion is included respectively. ($R_2$Ti$_2$O$_{7}$ compounds with $R=$ La, Pr, and Nd show qualitatively the same behavior).}\label{fig7}
\end{figure}
In order to understand better the ferroelectric symmetry breaking from the high-symmetry $P2_1/m$ parent structure to the low-symmetry polar $P2_1$ structure in the rare-earth titanates, we employed  symmetry-mode analysis using the Isotropy software from the Bilbao Crystallographic server \cite{orobengoa,perez}.
 We found that the symmetry-lowering distortion consists of displacements from two zone-center irreducible representations (irreps) of the $P2_1/m$ parent group, the $\Gamma1^+$ and $\Gamma1^-$ modes. Table \ref{modes} gives the isotropy subgroup and the amplitude percentage of the total distortion amplitude for each mode, and their corresponding atomic displacements are shown in Fig.~\ref{fig7}. Note that the symmetries and percentage contributions of these modes are very similar to those identified for the ferroelectric transition in the BaMF$_4$ family \cite{ederer}.
\begin{table}
\caption{\label{modes}Symmetry modes linking the $P2_1/m$ and $P2_1$ phases of the titanate Lichtenberg phases. The distortion amplitudes shown are for the case of CeTi$_2$O$_7$.}
\begin{tabular}{ccc}
\hline
\hline
Irrep &  Isotropy & Amplitude \% \\
	& Subgroup &   \\
\hline
$\Gamma1^+$ & $P2_1/m$ 	& 14.7 \\
$\Gamma1^-$ & $P2_1$ 	& 85.3  \\
\hline 
\hline
\end{tabular}
\end{table}
\begin{table*}
\caption{\label{dip-pol}Polarization quanta ($P_{q}$), and spontaneous polarizations ($P_{S}$) obtained using the Berry-phase method and from formal charges (Eq.(\ref{formalcharges})), for $R_2$Ti$_2$O$_7$ with $R=$ La, Pr, and Nd. Results from this work are labeled by [*]. All values are reported for the $P2_1/m\rightarrow P2_1$ transformation, except for those from Ref. \cite{lopez}, which are obtained for the  $Cmcm\rightarrow Cmc2_1$ transformation. Available literature experimental values, discussed in the Introduction, are also given for comparison.}
\begin{ruledtabular}
\begin{tabular}{lllccccc}
\multicolumn{4}{r}{} &  \multicolumn{1}{c} { Berry Phase } & \multicolumn{1}{c} { Formal charges } &  \multicolumn{2}{c} { Experimental }\\
\cline{5-8}
$R_2$Ti$_2$O$_7$ & Method &Ref. & $P_{q}$  ($\mu$C/cm$^2$) &  $P_{S}$ ($\mu$C/cm$^2$) & $P_{S}$ ($\mu$C/cm$^2$) &   $P_{S}$ ($\mu$C/cm$^2$)  & Ref.\\
\hline
La$_2$Ti$_2$O$_7$ & PBEsol & [*] 	& 15.77 &  18.20 & 14.44 & 5.0, 4.4 &  \cite{nana}, \cite{2002Kim} \\
				 & PW & \cite{bruyer} & 15.60 &  &7.72 & \\
				 & LDA & \cite{lopez} &16.02 & & 29.0 & \\ 
\\
Ce$_2$Ti$_2$O$_7$  &  PBEsol & [*] & 15.71  &  18.58  &  14.22 & 4.19 &  \cite{2008Kim}\\
\\

Pr$_2$Ti$_2$O$_7$ &  PBEsol & [*] & 15.80 & 18.17 & 14.18  & 7.0, 0.017 & \cite{2013Sun}, \cite{2015Patwe} \\
				 & PBE & \cite{2015Patwe} & 15.42 &  & 8.3 & \\
\\
				 
Nd$_2$Ti$_2$O$_7$ & PBEsol & [*] & 15.88 &  17.77  & 14.16 & 9.0, 1.8 &  \cite{kimura},  \cite{2002Kim} \\
				 & PW & \cite{bruyer} & 16.10 &  &7.42 & \\
\end{tabular}
\end{ruledtabular}
\end{table*}
\begin{figure}
\includegraphics[width=0.4\textwidth]{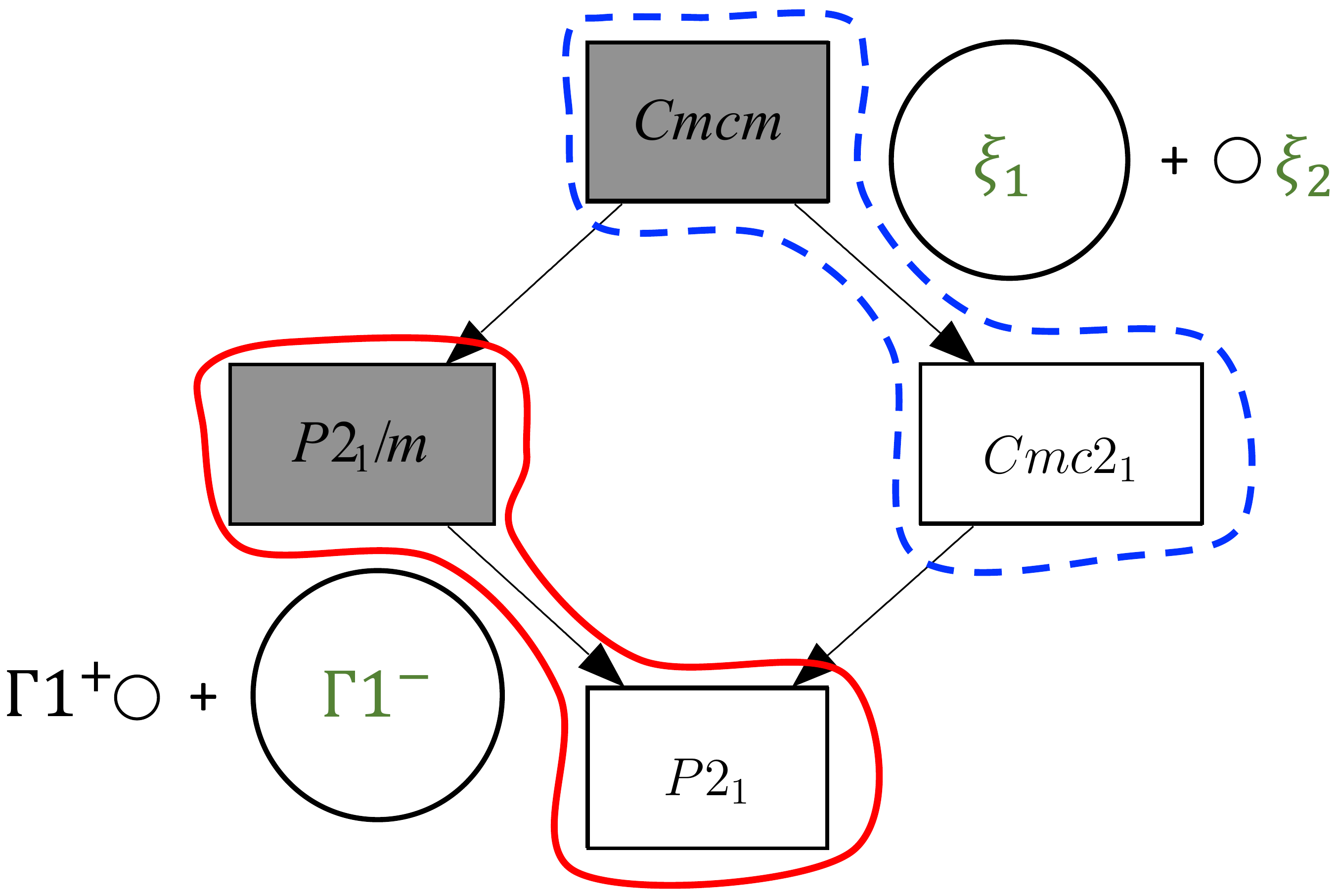}
\caption{\label{fig08} Mode decomposition in the centrosymmetric-to-polar transformations showing the pathway studied in this work (enclosed by the red solid line and modes: $\Gamma1^+$ and $\Gamma1^-$), as well as that in Ref.~\cite{lopez} (enclosed by the dashed blue line and modes: $\xi_1$ and $\xi_2$). Soft modes are shown in green. The area of the circles indicates the percentage contribution of the mode to the low-symmetry phase.} 
\end{figure}
\begin{figure}
\includegraphics[width=0.5\textwidth]{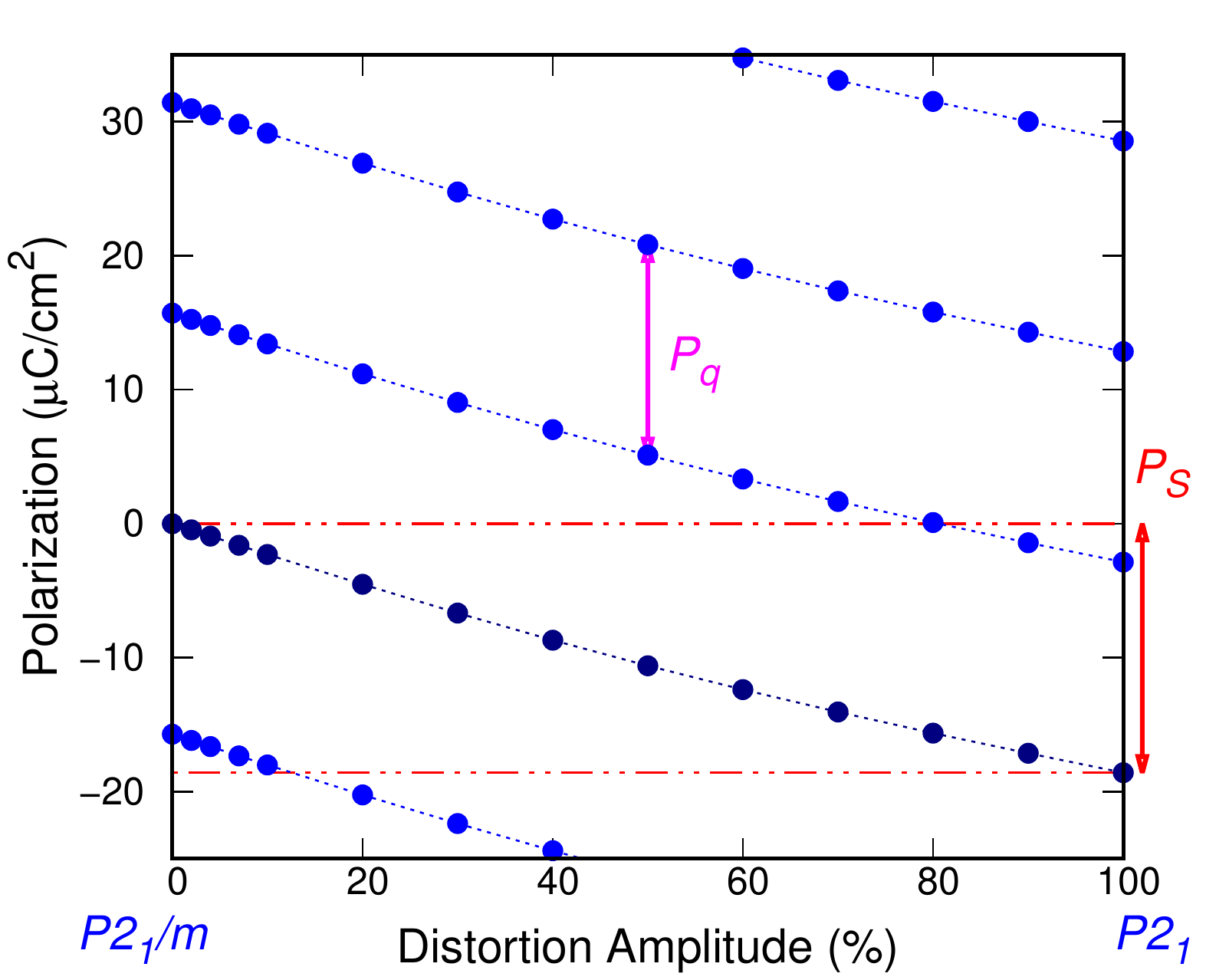}
\caption{\label{fig09} Polarization branches calculated as a function of distortion from the centrosymmetric parent $P2_1/m$ phase to the ferroelectric $P2_1$ phase of Ce$_2$Ti$_2$O$_7$.  The polarization branches are separated by the polarization quantum $P_{q}=15.71$~ $\mu$C/cm$^2$.}
\end{figure}
\begin{figure}
\includegraphics[width=0.5\textwidth]{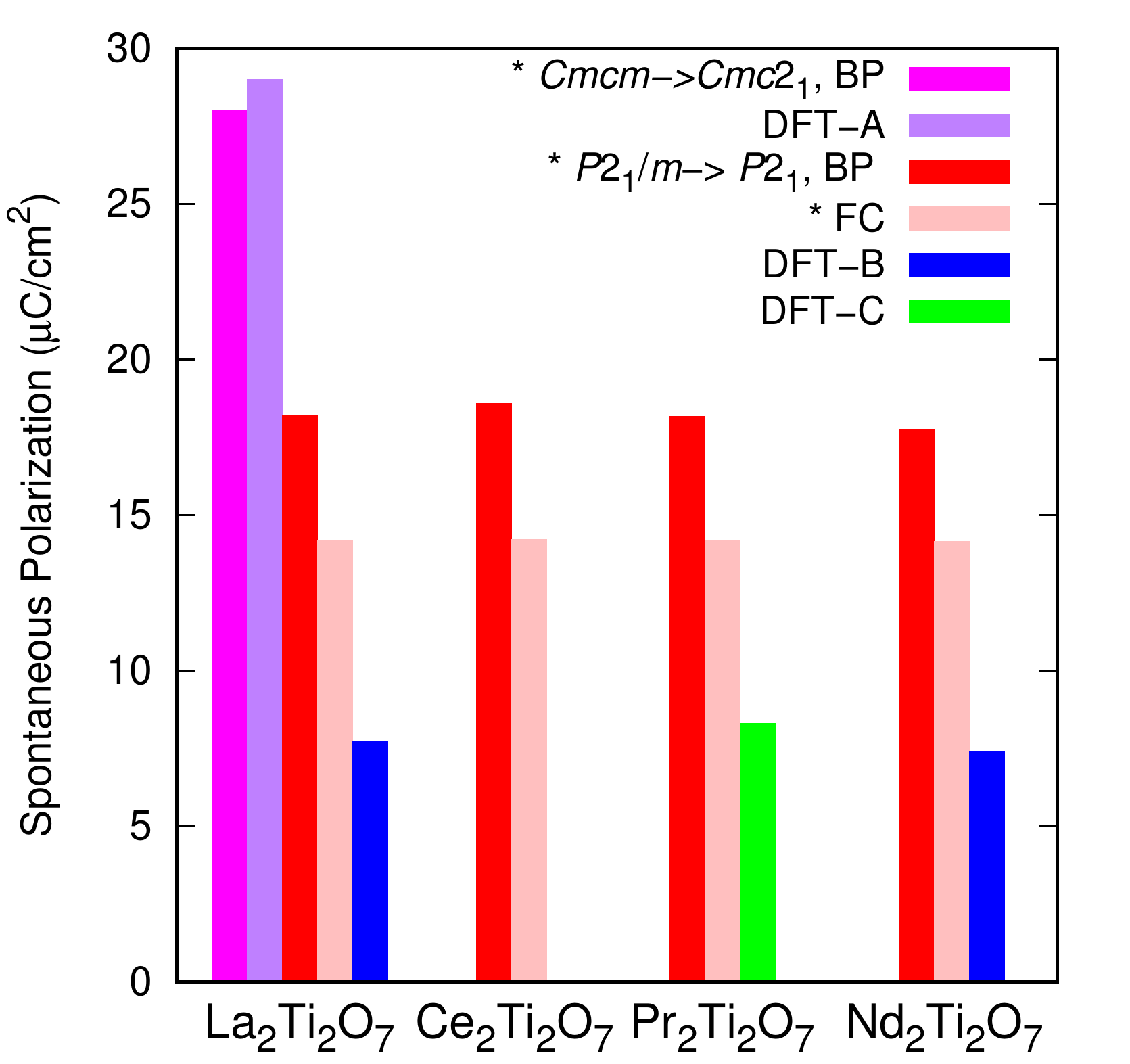}
\caption{\label{fig10} Calculated spontaneous polarizations using the Berry-phase (BP) method and Eq. (1) with formal charges (FC) for the $R_2$Ti$_2$O$_7$ materials with $R=$ La, Ce, Pr, and Nd. We compare our results labeled by * with other DFT calculations for LaTO, PrTO and NdTO from Refs. A ($Cmcm\rightarrow Cmc2_1$): \cite{lopez}, B: \cite{bruyer},  and C: \cite{2015Patwe}.} 
\end{figure}
The $\Gamma1^-$ mode, which consists of rotations and tiltings of the TiO$_6$  octahedra combined with displacements of the $R$ ions along the $b$ direction, is polar and lowers the symmetry to $P2_1$. It also lowers the energy from the parent $P2_1/m$ structure (Fig. \ref{fig7}(b) blue triangles), confirming that it is the primary order parameter for the transition, and indicating that Ce$_2$Ti$_2$O$_7$ is a proper ferroelectric. 
In contrast, the $\Gamma_1^+$ distortion  is a non-polar mode, consisting mainly of Ce-ion displacements in the $c$ direction (Fig. \ref{fig7}a) in such a way that it is symmetry conserving. In Fig. \ref{fig7}(b) we show the evolution of the energy  per formula unit as this mode is frozen in from the high-symmetry structure, and we see that it is not energy lowering and therefore not the primary order parameter for the transition. We also show in Fig. \ref{fig7}(b) the change in energy  per formula unit as one deforms the non-polar parent $P2_1/m$ structure to the ferroelectric $P2_1$ phase considering the total distortion amplitude, with 100\% distortion amplitude corresponding to the distortion value required to reach the polar $P2_1$ ground state.
We see that, while the $\Gamma1^+$ mode contributes only $\sim$15 \% of the total distortion amplitude, it approximately doubles the amount of energy lowering associated with the formation of the ground state compared with the $\Gamma1^-$ distortion alone. The ground-state structure, with both $\Gamma1^+$ and $\Gamma1^-$ distortions, is shown in the right panel of Fig.~\ref{fig7}(a). 
 The behavior of the $P2_1/m\rightarrow P2_1$ transformation is analogous to that reported in BaMF$_4$, in which the primary polar mode accounts for the majority of the total distortion, with a secondary non-polar mode also contributing to the ground-state structure \cite{ederer}.

To determine the magnitude of the spontaneous polarization, $P_S$, which occurs along the $b$ axis, we calculated the differences in polarization between the ferroelectric $P2_1$, and the centrosymmetric $P2_1/m$ phases for the $R_2$Ti$_2$O$_7$ series. Particular attention was paid in the mapping of the Berry-phase values onto the same branch of the polarization lattice, by calculating the polarization for a large number of intermediate structures along the deformation path between the $P2_1/m$  and $P2_1$ structures. Fig. \ref{fig09} shows the calculated polarization branches as a function of the norm of the distortion vectors (distortion amplitude) from the high-symmetry structure (0\%-distortion amplitude) to the ground state low-symmetry phase (100\%-distortion amplitude) for Ce$_2$Ti$_2$O$_7$ (for $R=$ La, Pr and Nd, the polarization branches, not shown here, are similar with the only difference being the polarization magnitude). 
The branches in Fig. \ref{fig09} are separated by the polarization quantum  $P_{q}=\frac{1}{V}eR_b$, where $V$ is the volume of the unit cell, $e$ is the electron charge, and $R_b$ is the $b$ lattice parameter. (We use the expression for spin-polarized systems to facilitate copmarison with magnetic Lichtenberg phases.) Furthermore, we also computed $P_S$ by multiplying the atomic displacements $\Delta X_{i,j}$ (where $i$ labels the atom and $j$ the direction of displacement) in the $P2_1/m\rightarrow P2_1$ transformation, multiplied by the formal charges, $Z=3$ for $R^{3+}$, $Z=4$ for Ti$^{4+}$, and $Z=-2$ for O$^{2-}$), that is,
\begin{equation}
P_{S,j}=\frac{1}{V}\sum{Z_{i}\Delta X_{i,j}} \quad . \label{formalcharges}
\end{equation}
Table III gives our Berry-phase and formal charge (from Eq. (1)) results for $R_2$Ti$_2$O$_7$ materials with $R=$ La, Ce, Pr, and Nd. For all materials, the Berry-phase polarizations are around 18 $\mu$C cm$^{-2}$, with the formal charge values slightly lower at around 14 $\mu$C cm$^{-2}$.
We see that multiplying the displacements of the ions with their formal ionic charges yields spontaneous polarizations that are $\sim$80\% of the Berry phase values. This close agreement is consistent with the geometric mechanism for ferroelectricity, in which the polarization arises from displacements of the ions without substantial electronic rehybridization \cite{ederer, 2004vanAken}. Our computed Born effective charge tensors $\mathbf{Z}^*$ (see Appendix, Tables IV-VII) indicate that, while the $\mathbf{Z}^*$s are close to their formal values for the majority of the ions, some Ti and O atoms have anomalously large effective charges of up to almost 8 for nominally Ti$^{4+}$ and up to $-5.5$ for nominally O$^{2-}$. These values are comparable to the anomalous Born effective charges in the prototypical conventional ferroelectrics BaTiO$_3$ and PbTiO$_3$ and indicate substantial rehybridization on displacement. These large effective charges correspond, however, to displacements along the $a$ direction in which the materials have infinite chains of TiO$_6$ octahedra, Figs. 1(b) and 4, like those in the conventional perovskite structure. They do not contribute to the net polarization because the displacements along the $a$ direction occur in an antipolar arrangement. 

We see that our calculated $P_S$ values for the $P2_1/m\rightarrow P2_1$ transformation are slightly more than double those reported in earlier DFT calcuations. 
This is not a result of our use of the PBEsol functional; our test calculations using PBE in fact gave larger $P_S$ values (not shown here) due to PBE's overestimation of the lattice 
parameters of the $P2_1/m$  and $P2_1$ structures (Table I), and in consequence the distortion amplitude between the two phases. 
We notice however, that, since $b<a$ and $b<<c$ (Table I), the polarization quantum $P_{q}$ along the $b$ axis is smaller than the spontaneous polarization $P_{S}$. This feature could be a source of problems in correctly connecting the polarization lattice points to obtain the polarization branches if insufficient intermediate distortions are taken, and might explain the lower-than-expected calculated values reported in the literature. Unfortunately no polarization lattices equivalent to those shown in Fig.~\ref{fig09} were presented in the earlier works \cite{bruyer,2015Patwe}. 
\begin{figure}
\includegraphics[width=0.5\textwidth]{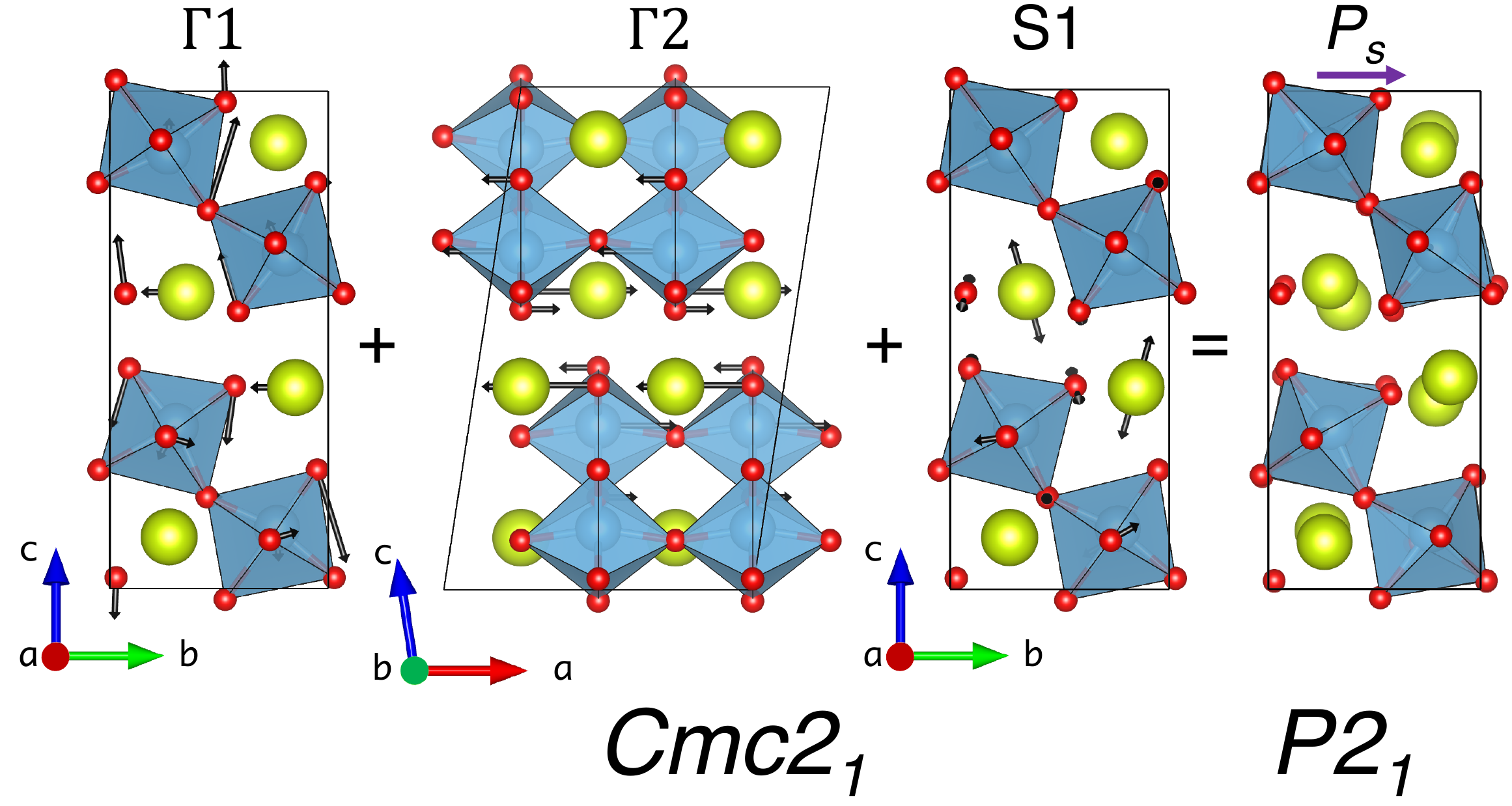}
\caption{\label{fig11}Representation of the mode decomposition in the $Cmc2_1\rightarrow P2_1$ transformation of $R_2$Ti$_2$O$_7$. The arrows indicate the magnitude (doubled in length for visualization purposes) and direction of the atomic displacements, starting from the high-symmetry structure, in each mode.} 
\end{figure}
Finally, we comment on the difference between our calculated polarization value for the $P2_1/m \rightarrow P2_1$ transition in LaTO (18.2 $\mu$C/cm$^2$) and the larger value obtained in the earlier study of the $Cmcm\rightarrow Cmc2_1$ transition in the same material  (29.0 $\mu$C/cm$^2$) \cite{lopez}. 
The mode decomposition and energy profile results reported for the $Cmcm\rightarrow Cmc2_1$ transition also indicated the contribution of two modes, although in contrast to the $P2_1/m\rightarrow P2_1$ transition, in the former case both are polar and both are soft, with one being marginally unstable; we indicate the relative contributions of each mode to the final structure in Fig.~\ref{fig08}.
We begin by calculating the polarization for the $Cmcm\rightarrow Cmc2_1$ transition of LaTO using the computational parameters of Ref.~[\onlinecite{lopez}], and find that our calculation (shown by the magenta line in Fig.~\ref{fig10}) indeed closely reproduces the literature value (purple line). As a double check, we calculated the polarizations of the structures in all four  space groups from summing the formal charges multiplied by the displacements of the atoms from their centrosymmetric positions taking the same lattice parameters for all structures: For $P2_1/m$ and $Cmcm$ we of course obtained values of zero by definition, for $P2_1$ we obtained 14.4~$\mu$C/cm$^2$, and for $Cmc2_1$ 19.5~$\mu$C/cm$^2$, both underestimating the full Berry phase values as discussed above; the polarization difference obtained by displacing the atoms along a pathway connecting the $P2_1\rightarrow Cmc2_1$ phases is consistent at 5.1 $\mu$C/cm$^2$. This leads us to the unusual conclusion that the higher-symmetry $Cmc2_1$ phase has a larger spontaneous polarization than the lower symmetry $P2_1$ phase, and that if the transition follows the pathway $Cmcm\rightarrow Cmc2_1 \rightarrow P2_1$ on reducing temperature, the system will first show an increase in polarization from zero, then a decrease to a smaller non-zero value at the two successive phase transitions. The distortion modes that link the $Cmc2_1$ and $P2_1$ structures are shown in Fig.~\ref{fig11}.
%
\section{Summary and Conclusions}
In summary, we investigated the structural and ferroelectric properties of the family of perovskite-related layered rare-earth titanate Lichtenberg phases, with chemical composition $R_2$Ti$_2$O$_7$ for $R=$ La, Ce, Pr, and Nd using first-principles calculations. We studied the mechanism of the ferroelectric distortion between the lowest energy polar phase, belonging to the monoclinic space group $P2_1$, and its parent high-symmetry non-polar $P2_1/m$ structure. We found that the energy lowerings associated with the ferroelectric transitions are consistent with their high Curie temperatures compared to those of conventional ferroelectrics such as BaTiO$_3$. Interestingly, while the energy lowering between the non-polar and polar structures increases across the La - Ce - Pr - Nd series, consistent with the smaller rare-earth cation allowing larger structural distortions, the polarization is approximately constant across the series with a value of around 18 $\mu$C/cm$^2$. 
We found that the ferroelectric transition is driven by a polar distortion of $\Gamma1^-$ symmetry consisting of tilts and rotations of the TiO$_6$ octahedra combined with small $R^{3+}$ cation displacements; a secondary distortion  of $\Gamma1^+$ also contributes to the ground-state structure. This mode decomposition result is similar to that identified in the ferroelectric transformation of the Ba$M$F$_4$ family ($M=$Mn, Fe, Co, Ni) \cite{ederer}, although the slightly anomalous Born effective charges indicate a higher degree of chemical rehybridization across the transition in this case. 

We compared our results to an earlier literature study of the higher-energy $Cmcm\rightarrow Cmc2_1$ transition, and noted that the higher energy, higher symmetry $Cmc2_1$ has a higher polarization than the lower energy, lower symmetry $P2_1$ phase. This suggests a possible unusual {\it decrease} in ferroelectric polarization on cooling if this pathway is followed. We hope that our results motivate additional experimental studies of this fascinating material class, in particular to resolve the sequence of phases that occur during the ferroelectric phase transition, and to realize the substantial polarization values that have not yet been achieved experimentally.

\begin{acknowledgments}
This work was supported by the ETH-Z\"urich and the Deutsches GeoForschungZentrum, Postdam. Calculations were performed on the ETH-Z\"urich Brutus cluster.  The authors also gratefully acknowledge the Gauss Centre for Supercomputing e.V. (www.gauss-centre.eu) for funding this project by providing computing time through the John von Neumann Institute for Computing (NIC) on the GCS Supercomputer JUWELS at Jülich Supercomputing Centre (JSC) under project HPO24.
\end{acknowledgments}
\bibliography{Re2Ti2O7_paper}

\providecommand{\noopsort}[1]{}\providecommand{\singleletter}[1]{#1}%
\begin{thebibliography}{52}%
\makeatletter
\providecommand \@ifxundefined [1]{%
 \@ifx{#1\undefined}
}%
\providecommand \@ifnum [1]{%
 \ifnum #1\expandafter \@firstoftwo
 \else \expandafter \@secondoftwo
 \fi
}%
\providecommand \@ifx [1]{%
 \ifx #1\expandafter \@firstoftwo
 \else \expandafter \@secondoftwo
 \fi
}%
\providecommand \natexlab [1]{#1}%
\providecommand \enquote  [1]{``#1''}%
\providecommand \bibnamefont  [1]{#1}%
\providecommand \bibfnamefont [1]{#1}%
\providecommand \citenamefont [1]{#1}%
\providecommand \href@noop [0]{\@secondoftwo}%
\providecommand \href [0]{\begingroup \@sanitize@url \@href}%
\providecommand \@href[1]{\@@startlink{#1}\@@href}%
\providecommand \@@href[1]{\endgroup#1\@@endlink}%
\providecommand \@sanitize@url [0]{\catcode `\\12\catcode `\$12\catcode
  `\&12\catcode `\#12\catcode `\^12\catcode `\_12\catcode `\%12\relax}%
\providecommand \@@startlink[1]{}%
\providecommand \@@endlink[0]{}%
\providecommand \url  [0]{\begingroup\@sanitize@url \@url }%
\providecommand \@url [1]{\endgroup\@href {#1}{\urlprefix }}%
\providecommand \urlprefix  [0]{URL }%
\providecommand \Eprint [0]{\href }%
\providecommand \doibase [0]{http://dx.doi.org/}%
\providecommand \selectlanguage [0]{\@gobble}%
\providecommand \bibinfo  [0]{\@secondoftwo}%
\providecommand \bibfield  [0]{\@secondoftwo}%
\providecommand \translation [1]{[#1]}%
\providecommand \BibitemOpen [0]{}%
\providecommand \bibitemStop [0]{}%
\providecommand \bibitemNoStop [0]{.\EOS\space}%
\providecommand \EOS [0]{\spacefactor3000\relax}%
\providecommand \BibitemShut  [1]{\csname bibitem#1\endcsname}%
\let\auto@bib@innerbib\@empty
\bibitem [{\citenamefont {Lichtenberg}\ \emph {et~al.}(2008)\citenamefont
  {Lichtenberg}, \citenamefont {Herrnberger},\ and\ \citenamefont
  {Wiedemann}}]{frank}%
  \BibitemOpen
  \bibfield  {author} {\bibinfo {author} {\bibfnamefont {F.}~\bibnamefont
  {Lichtenberg}}, \bibinfo {author} {\bibfnamefont {A.}~\bibnamefont
  {Herrnberger}}, \ and\ \bibinfo {author} {\bibfnamefont {K.}~\bibnamefont
  {Wiedemann}},\ }\href@noop {} {\bibfield  {journal} {\bibinfo  {journal}
  {Prog.\ Sol.\ Stat.\ Chem.}\ }\textbf {\bibinfo {volume} {36}},\ \bibinfo
  {pages} {253} (\bibinfo {year} {2008})}\BibitemShut {NoStop}%
\bibitem [{\citenamefont {Nanot}\ \emph {et~al.}(1981)\citenamefont {Nanot},
  \citenamefont {Queyroux}, \citenamefont {Gilles},\ and\ \citenamefont
  {Portier}}]{1981Nanot}%
  \BibitemOpen
  \bibfield  {author} {\bibinfo {author} {\bibfnamefont {M.}~\bibnamefont
  {Nanot}}, \bibinfo {author} {\bibfnamefont {F.}~\bibnamefont {Queyroux}},
  \bibinfo {author} {\bibfnamefont {J.-C.}\ \bibnamefont {Gilles}}, \ and\
  \bibinfo {author} {\bibfnamefont {R.}~\bibnamefont {Portier}},\ }\href@noop
  {} {\bibfield  {journal} {\bibinfo  {journal} {J. Solid State Chem.}\
  }\textbf {\bibinfo {volume} {38}},\ \bibinfo {pages} {74} (\bibinfo {year}
  {1981})}\BibitemShut {NoStop}%
\bibitem [{\citenamefont {Saha}\ \emph {et~al.}(2011)\citenamefont {Saha},
  \citenamefont {Prusty}, \citenamefont {Singh}, \citenamefont
  {Suryanarayanan}, \citenamefont {Revcolevschi},\ and\ \citenamefont
  {Sood}}]{saha}%
  \BibitemOpen
  \bibfield  {author} {\bibinfo {author} {\bibfnamefont {S.}~\bibnamefont
  {Saha}}, \bibinfo {author} {\bibfnamefont {S.}~\bibnamefont {Prusty}},
  \bibinfo {author} {\bibfnamefont {S.}~\bibnamefont {Singh}}, \bibinfo
  {author} {\bibfnamefont {R.}~\bibnamefont {Suryanarayanan}}, \bibinfo
  {author} {\bibfnamefont {A.}~\bibnamefont {Revcolevschi}}, \ and\ \bibinfo
  {author} {\bibfnamefont {A.~K.}\ \bibnamefont {Sood}},\ }\href@noop {}
  {\bibfield  {journal} {\bibinfo  {journal} {Sol. Stat. Chem.}\ }\textbf
  {\bibinfo {volume} {184}},\ \bibinfo {pages} {2204} (\bibinfo {year}
  {2011})}\BibitemShut {NoStop}%
\bibitem [{\citenamefont {Subramanian}\ \emph {et~al.}(1983)\citenamefont
  {Subramanian}, \citenamefont {Aravamudan},\ and\ \citenamefont
  {Rao}}]{pyrochlore}%
  \BibitemOpen
  \bibfield  {author} {\bibinfo {author} {\bibfnamefont {M.}~\bibnamefont
  {Subramanian}}, \bibinfo {author} {\bibfnamefont {G.}~\bibnamefont
  {Aravamudan}}, \ and\ \bibinfo {author} {\bibfnamefont {G.~V.~S.}\
  \bibnamefont {Rao}},\ }\href@noop {} {\bibfield  {journal} {\bibinfo
  {journal} {Prog. in Sol. Stat. Chem.}\ }\textbf {\bibinfo {volume} {15}},\
  \bibinfo {pages} {55} (\bibinfo {year} {1983})}\BibitemShut {NoStop}%
\bibitem [{\citenamefont {Titov}\ \emph {et~al.}(1987)\citenamefont {Titov},
  \citenamefont {Sych}, \citenamefont {Mel?nik},\ and\ \citenamefont
  {Bondarenko}}]{1987Titov}%
  \BibitemOpen
  \bibfield  {author} {\bibinfo {author} {\bibfnamefont {Y.~A.}\ \bibnamefont
  {Titov}}, \bibinfo {author} {\bibfnamefont {A.~M.}\ \bibnamefont {Sych}},
  \bibinfo {author} {\bibfnamefont {V.~M.}\ \bibnamefont {Mel?nik}}, \ and\
  \bibinfo {author} {\bibfnamefont {T.~N.}\ \bibnamefont {Bondarenko}},\
  }\href@noop {} {\bibfield  {journal} {\bibinfo  {journal} {Russ. J. Inorg.
  Chem.}\ }\textbf {\bibinfo {volume} {32}},\ \bibinfo {pages} {3} (\bibinfo
  {year} {1987})}\BibitemShut {NoStop}%
\bibitem [{\citenamefont {Shao}\ \emph
  {et~al.}(2012{\natexlab{a}})\citenamefont {Shao}, \citenamefont {Saitzek},
  \citenamefont {Ferri}, \citenamefont {Rguiti}, \citenamefont {Dupont},
  \citenamefont {Roussel},\ and\ \citenamefont {Desfeux}}]{2012Shao}%
  \BibitemOpen
  \bibfield  {author} {\bibinfo {author} {\bibfnamefont {Z.}~\bibnamefont
  {Shao}}, \bibinfo {author} {\bibfnamefont {S.}~\bibnamefont {Saitzek}},
  \bibinfo {author} {\bibfnamefont {A.}~\bibnamefont {Ferri}}, \bibinfo
  {author} {\bibfnamefont {M.}~\bibnamefont {Rguiti}}, \bibinfo {author}
  {\bibfnamefont {L.}~\bibnamefont {Dupont}}, \bibinfo {author} {\bibfnamefont
  {P.}~\bibnamefont {Roussel}}, \ and\ \bibinfo {author} {\bibfnamefont
  {R.}~\bibnamefont {Desfeux}},\ }\href@noop {} {\bibfield  {journal} {\bibinfo
   {journal} {J. Mater. Chem.}\ }\textbf {\bibinfo {volume} {22}},\ \bibinfo
  {pages} {9806} (\bibinfo {year} {2012}{\natexlab{a}})}\BibitemShut {NoStop}%
\bibitem [{\citenamefont {Shao}\ \emph
  {et~al.}(2012{\natexlab{b}})\citenamefont {Shao}, \citenamefont {Saitzek},
  \citenamefont {Roussel},\ and\ \citenamefont {Desfeux}}]{2012Shaob}%
  \BibitemOpen
  \bibfield  {author} {\bibinfo {author} {\bibfnamefont {Z.}~\bibnamefont
  {Shao}}, \bibinfo {author} {\bibfnamefont {S.}~\bibnamefont {Saitzek}},
  \bibinfo {author} {\bibfnamefont {P.}~\bibnamefont {Roussel}}, \ and\
  \bibinfo {author} {\bibfnamefont {R.}~\bibnamefont {Desfeux}},\ }\href@noop
  {} {\bibfield  {journal} {\bibinfo  {journal} {J. Mater. Chem.}\ }\textbf
  {\bibinfo {volume} {22}},\ \bibinfo {pages} {24894} (\bibinfo {year}
  {2012}{\natexlab{b}})}\BibitemShut {NoStop}%
\bibitem [{\citenamefont {Damjanovic}(1998)}]{damjanovic}%
  \BibitemOpen
  \bibfield  {author} {\bibinfo {author} {\bibfnamefont {D.}~\bibnamefont
  {Damjanovic}},\ }\href@noop {} {\bibfield  {journal} {\bibinfo  {journal}
  {Curr. Opin. in Sol. Stat. and Mat. Sci.}\ }\textbf {\bibinfo {volume} {3}},\
  \bibinfo {pages} {469} (\bibinfo {year} {1998})}\BibitemShut {NoStop}%
\bibitem [{\citenamefont {Turner}\ \emph {et~al.}(1994)\citenamefont {Turner},
  \citenamefont {Fuierer}, \citenamefont {Newnham},\ and\ \citenamefont
  {Shrout}}]{turner}%
  \BibitemOpen
  \bibfield  {author} {\bibinfo {author} {\bibfnamefont {R.~C.}\ \bibnamefont
  {Turner}}, \bibinfo {author} {\bibfnamefont {P.~A.}\ \bibnamefont {Fuierer}},
  \bibinfo {author} {\bibfnamefont {R.~E.}\ \bibnamefont {Newnham}}, \ and\
  \bibinfo {author} {\bibfnamefont {T.~R.}\ \bibnamefont {Shrout}},\
  }\href@noop {} {\bibfield  {journal} {\bibinfo  {journal} {Appl. Acoust.}\
  }\textbf {\bibinfo {volume} {41}},\ \bibinfo {pages} {299} (\bibinfo {year}
  {1994})}\BibitemShut {NoStop}%
\bibitem [{\citenamefont {Atuchin}\ \emph {et~al.}(2012)\citenamefont
  {Atuchin}, \citenamefont {Gavrilova}, \citenamefont {Grivel}, \citenamefont
  {Kesler},\ and\ \citenamefont {Troitskaia}}]{atuchin}%
  \BibitemOpen
  \bibfield  {author} {\bibinfo {author} {\bibfnamefont {V.~V.}\ \bibnamefont
  {Atuchin}}, \bibinfo {author} {\bibfnamefont {T.~A.}\ \bibnamefont
  {Gavrilova}}, \bibinfo {author} {\bibfnamefont {J.~C.}\ \bibnamefont
  {Grivel}}, \bibinfo {author} {\bibfnamefont {V.~G.}\ \bibnamefont {Kesler}},
  \ and\ \bibinfo {author} {\bibfnamefont {I.~B.}\ \bibnamefont {Troitskaia}},\
  }\href@noop {} {\bibfield  {journal} {\bibinfo  {journal} {J. Sol. Stat.
  Chem.}\ }\textbf {\bibinfo {volume} {195}},\ \bibinfo {pages} {125} (\bibinfo
  {year} {2012})}\BibitemShut {NoStop}%
\bibitem [{\citenamefont {Hwang}\ \emph {et~al.}(2003)\citenamefont {Hwang},
  \citenamefont {Lee}, \citenamefont {Li},\ and\ \citenamefont {Oh}}]{hwang}%
  \BibitemOpen
  \bibfield  {author} {\bibinfo {author} {\bibfnamefont {D.~W.}\ \bibnamefont
  {Hwang}}, \bibinfo {author} {\bibfnamefont {J.~S.}\ \bibnamefont {Lee}},
  \bibinfo {author} {\bibfnamefont {W.}~\bibnamefont {Li}}, \ and\ \bibinfo
  {author} {\bibfnamefont {S.~H.}\ \bibnamefont {Oh}},\ }\href@noop {}
  {\bibfield  {journal} {\bibinfo  {journal} {J. Phys. Chem. B}\ }\textbf
  {\bibinfo {volume} {107}},\ \bibinfo {pages} {4963} (\bibinfo {year}
  {2003})}\BibitemShut {NoStop}%
\bibitem [{\citenamefont {Nanamatsu}\ \emph {et~al.}(1974)\citenamefont
  {Nanamatsu}, \citenamefont {Kimura}, \citenamefont {Koi}, \citenamefont
  {Matsushita},\ and\ \citenamefont {Yamada}}]{nana}%
  \BibitemOpen
  \bibfield  {author} {\bibinfo {author} {\bibfnamefont {S.}~\bibnamefont
  {Nanamatsu}}, \bibinfo {author} {\bibfnamefont {M.}~\bibnamefont {Kimura}},
  \bibinfo {author} {\bibfnamefont {K.}~\bibnamefont {Koi}}, \bibinfo {author}
  {\bibfnamefont {S.}~\bibnamefont {Matsushita}}, \ and\ \bibinfo {author}
  {\bibfnamefont {N.}~\bibnamefont {Yamada}},\ }\href@noop {} {\bibfield
  {journal} {\bibinfo  {journal} {Ferroelectrics}\ }\textbf {\bibinfo {volume}
  {8}},\ \bibinfo {pages} {511} (\bibinfo {year} {1974})}\BibitemShut {NoStop}%
\bibitem [{\citenamefont {Kimura}\ \emph {et~al.}(1974)\citenamefont {Kimura},
  \citenamefont {Nanamatsu}, \citenamefont {Kawamura},\ and\ \citenamefont
  {Matsushita}}]{kimura}%
  \BibitemOpen
  \bibfield  {author} {\bibinfo {author} {\bibfnamefont {M.}~\bibnamefont
  {Kimura}}, \bibinfo {author} {\bibfnamefont {S.}~\bibnamefont {Nanamatsu}},
  \bibinfo {author} {\bibfnamefont {T.}~\bibnamefont {Kawamura}}, \ and\
  \bibinfo {author} {\bibfnamefont {S.}~\bibnamefont {Matsushita}},\
  }\href@noop {} {\bibfield  {journal} {\bibinfo  {journal} {Jpn. J. Appl.
  Phys.}\ }\textbf {\bibinfo {volume} {13}},\ \bibinfo {pages} {1473} (\bibinfo
  {year} {1974})}\BibitemShut {NoStop}%
\bibitem [{\citenamefont {Kim}\ \emph {et~al.}(2008)\citenamefont {Kim},
  \citenamefont {Yang}, \citenamefont {Lee}, \citenamefont {Lee},\ and\
  \citenamefont {Park}}]{2008Kim}%
  \BibitemOpen
  \bibfield  {author} {\bibinfo {author} {\bibfnamefont {W.}~\bibnamefont
  {Kim}}, \bibinfo {author} {\bibfnamefont {J.}~\bibnamefont {Yang}}, \bibinfo
  {author} {\bibfnamefont {C.}~\bibnamefont {Lee}}, \bibinfo {author}
  {\bibfnamefont {H.}~\bibnamefont {Lee}}, \ and\ \bibinfo {author}
  {\bibfnamefont {H.}~\bibnamefont {Park}},\ }\href@noop {} {\bibfield
  {journal} {\bibinfo  {journal} {Thin Solid Films}\ }\textbf {\bibinfo
  {volume} {517}},\ \bibinfo {pages} {506} (\bibinfo {year}
  {2008})}\BibitemShut {NoStop}%
\bibitem [{\citenamefont {Sun}\ \emph {et~al.}(2013)\citenamefont {Sun},
  \citenamefont {Ju}, \citenamefont {Qin}, \citenamefont {Zhao}, \citenamefont
  {Su},\ and\ \citenamefont {Hu}}]{2013Sun}%
  \BibitemOpen
  \bibfield  {author} {\bibinfo {author} {\bibfnamefont {L.}~\bibnamefont
  {Sun}}, \bibinfo {author} {\bibfnamefont {L.}~\bibnamefont {Ju}}, \bibinfo
  {author} {\bibfnamefont {H.}~\bibnamefont {Qin}}, \bibinfo {author}
  {\bibfnamefont {M.}~\bibnamefont {Zhao}}, \bibinfo {author} {\bibfnamefont
  {W.}~\bibnamefont {Su}}, \ and\ \bibinfo {author} {\bibfnamefont
  {J.}~\bibnamefont {Hu}},\ }\href@noop {} {\bibfield  {journal} {\bibinfo
  {journal} {Physica B}\ }\textbf {\bibinfo {volume} {431}},\ \bibinfo {pages}
  {49} (\bibinfo {year} {2013})}\BibitemShut {NoStop}%
\bibitem [{\citenamefont {Ishizawa}\ \emph {et~al.}(1982)\citenamefont
  {Ishizawa}, \citenamefont {Marumo}, \citenamefont {Iwai}, \citenamefont
  {Kimura},\ and\ \citenamefont {Kawamura}}]{ishizawa2}%
  \BibitemOpen
  \bibfield  {author} {\bibinfo {author} {\bibfnamefont {N.}~\bibnamefont
  {Ishizawa}}, \bibinfo {author} {\bibfnamefont {F.}~\bibnamefont {Marumo}},
  \bibinfo {author} {\bibfnamefont {S.}~\bibnamefont {Iwai}}, \bibinfo {author}
  {\bibfnamefont {M.}~\bibnamefont {Kimura}}, \ and\ \bibinfo {author}
  {\bibfnamefont {T.}~\bibnamefont {Kawamura}},\ }\href@noop {} {\bibfield
  {journal} {\bibinfo  {journal} {Acta Cryst.}\ }\textbf {\bibinfo {volume}
  {B38}},\ \bibinfo {pages} {368} (\bibinfo {year} {1982})}\BibitemShut
  {NoStop}%
\bibitem [{\citenamefont {Patwe}\ \emph {et~al.}(2015)\citenamefont {Patwe},
  \citenamefont {Katari}, \citenamefont {Salke}, \citenamefont {Deshpande},
  \citenamefont {Rao}, \citenamefont {Gupta}, \citenamefont {Mittal},
  \citenamefont {Achary},\ and\ \citenamefont {Tyagi}}]{2015Patwe}%
  \BibitemOpen
  \bibfield  {author} {\bibinfo {author} {\bibfnamefont {S.~J.}\ \bibnamefont
  {Patwe}}, \bibinfo {author} {\bibfnamefont {V.}~\bibnamefont {Katari}},
  \bibinfo {author} {\bibfnamefont {N.~P.}\ \bibnamefont {Salke}}, \bibinfo
  {author} {\bibfnamefont {S.~K.}\ \bibnamefont {Deshpande}}, \bibinfo {author}
  {\bibfnamefont {R.}~\bibnamefont {Rao}}, \bibinfo {author} {\bibfnamefont
  {M.~K.}\ \bibnamefont {Gupta}}, \bibinfo {author} {\bibfnamefont
  {R.}~\bibnamefont {Mittal}}, \bibinfo {author} {\bibfnamefont {S.~N.}\
  \bibnamefont {Achary}}, \ and\ \bibinfo {author} {\bibfnamefont {A.~K.}\
  \bibnamefont {Tyagi}},\ }\href@noop {} {\bibfield  {journal} {\bibinfo
  {journal} {J. Mater. Chem. C}\ }\textbf {\bibinfo {volume} {3}},\ \bibinfo
  {pages} {4570} (\bibinfo {year} {2015})}\BibitemShut {NoStop}%
\bibitem [{\citenamefont {Kim}\ \emph {et~al.}(2002)\citenamefont {Kim},
  \citenamefont {Ha}, \citenamefont {Yun},\ and\ \citenamefont
  {Park}}]{2002Kim}%
  \BibitemOpen
  \bibfield  {author} {\bibinfo {author} {\bibfnamefont {W.}~\bibnamefont
  {Kim}}, \bibinfo {author} {\bibfnamefont {S.}~\bibnamefont {Ha}}, \bibinfo
  {author} {\bibfnamefont {S.}~\bibnamefont {Yun}}, \ and\ \bibinfo {author}
  {\bibfnamefont {H.}~\bibnamefont {Park}},\ }\href@noop {} {\bibfield
  {journal} {\bibinfo  {journal} {Thin Solid Films}\ }\textbf {\bibinfo
  {volume} {420-421}},\ \bibinfo {pages} {575} (\bibinfo {year}
  {2002})}\BibitemShut {NoStop}%
\bibitem [{\citenamefont {Bayart}\ \emph {et~al.}(2013)\citenamefont {Bayart},
  \citenamefont {Saitzek}, \citenamefont {Chambrier}, \citenamefont {Shao},
  \citenamefont {Ferri}, \citenamefont {Huv{\'e}}, \citenamefont {Pouhet},
  \citenamefont {Tebano}, \citenamefont {Roussel},\ and\ \citenamefont
  {Desfeux}}]{2013Bayart}%
  \BibitemOpen
  \bibfield  {author} {\bibinfo {author} {\bibfnamefont {A.}~\bibnamefont
  {Bayart}}, \bibinfo {author} {\bibfnamefont {S.}~\bibnamefont {Saitzek}},
  \bibinfo {author} {\bibfnamefont {M.-H.}\ \bibnamefont {Chambrier}}, \bibinfo
  {author} {\bibfnamefont {Z.}~\bibnamefont {Shao}}, \bibinfo {author}
  {\bibfnamefont {A.}~\bibnamefont {Ferri}}, \bibinfo {author} {\bibfnamefont
  {M.}~\bibnamefont {Huv{\'e}}}, \bibinfo {author} {\bibfnamefont
  {R.}~\bibnamefont {Pouhet}}, \bibinfo {author} {\bibfnamefont
  {A.}~\bibnamefont {Tebano}}, \bibinfo {author} {\bibfnamefont
  {P.}~\bibnamefont {Roussel}}, \ and\ \bibinfo {author} {\bibfnamefont
  {R.}~\bibnamefont {Desfeux}},\ }\href@noop {} {\bibfield  {journal} {\bibinfo
   {journal} {Cryst. Eng. Comm.}\ }\textbf {\bibinfo {volume} {15}},\ \bibinfo
  {pages} {4341} (\bibinfo {year} {2013})}\BibitemShut {NoStop}%
\bibitem [{\citenamefont {Bayart}\ \emph {et~al.}(2014)\citenamefont {Bayart},
  \citenamefont {Saitzek}, \citenamefont {Ferri}, \citenamefont {Pouhet},
  \citenamefont {Chambrier}, \citenamefont {Roussel},\ and\ \citenamefont
  {Desfeux}}]{2014Bayart}%
  \BibitemOpen
  \bibfield  {author} {\bibinfo {author} {\bibfnamefont {A.}~\bibnamefont
  {Bayart}}, \bibinfo {author} {\bibfnamefont {S.}~\bibnamefont {Saitzek}},
  \bibinfo {author} {\bibfnamefont {A.}~\bibnamefont {Ferri}}, \bibinfo
  {author} {\bibfnamefont {R.}~\bibnamefont {Pouhet}}, \bibinfo {author}
  {\bibfnamefont {M.-H.}\ \bibnamefont {Chambrier}}, \bibinfo {author}
  {\bibfnamefont {P.}~\bibnamefont {Roussel}}, \ and\ \bibinfo {author}
  {\bibfnamefont {R.}~\bibnamefont {Desfeux}},\ }\href@noop {} {\bibfield
  {journal} {\bibinfo  {journal} {Thin Solid Films}\ }\textbf {\bibinfo
  {volume} {553}},\ \bibinfo {pages} {71 } (\bibinfo {year}
  {2014})}\BibitemShut {NoStop}%
\bibitem [{\citenamefont {Bayart}\ \emph {et~al.}(2016)\citenamefont {Bayart},
  \citenamefont {Shao}, \citenamefont {Ferri}, \citenamefont {Roussel},
  \citenamefont {Desfeux},\ and\ \citenamefont {Saitzek}}]{2016Bayart}%
  \BibitemOpen
  \bibfield  {author} {\bibinfo {author} {\bibfnamefont {A.}~\bibnamefont
  {Bayart}}, \bibinfo {author} {\bibfnamefont {Z.}~\bibnamefont {Shao}},
  \bibinfo {author} {\bibfnamefont {A.}~\bibnamefont {Ferri}}, \bibinfo
  {author} {\bibfnamefont {P.}~\bibnamefont {Roussel}}, \bibinfo {author}
  {\bibfnamefont {R.}~\bibnamefont {Desfeux}}, \ and\ \bibinfo {author}
  {\bibfnamefont {S.}~\bibnamefont {Saitzek}},\ }\href@noop {} {\bibfield
  {journal} {\bibinfo  {journal} {RSC Adv.}\ }\textbf {\bibinfo {volume} {6}},\
  \bibinfo {pages} {32994} (\bibinfo {year} {2016})}\BibitemShut {NoStop}%
\bibitem [{\citenamefont {L{\'{o}pez-P{\'{e}}rez and J.
  {\'{I}}{\~{n}}iguez}}(2011)}]{lopez}%
  \BibitemOpen
  \bibfield  {author} {\bibinfo {author} {\bibfnamefont {J.}~\bibnamefont
  {L{\'{o}pez-P{\'{e}}rez and J. {\'{I}}{\~{n}}iguez}}},\ }\href@noop {}
  {\bibfield  {journal} {\bibinfo  {journal} {Phys. Rev. B}\ }\textbf {\bibinfo
  {volume} {84}},\ \bibinfo {pages} {075121} (\bibinfo {year}
  {2011})}\BibitemShut {NoStop}%
\bibitem [{\citenamefont {Ivantchev}\ \emph {et~al.}(2000)\citenamefont
  {Ivantchev}, \citenamefont {Kroumova}, \citenamefont {Madariaga},
  \citenamefont {P{\'e}rez-Mato},\ and\ \citenamefont {Aroyo}}]{2000Ivantchev}%
  \BibitemOpen
  \bibfield  {author} {\bibinfo {author} {\bibfnamefont {S.}~\bibnamefont
  {Ivantchev}}, \bibinfo {author} {\bibfnamefont {E.}~\bibnamefont {Kroumova}},
  \bibinfo {author} {\bibfnamefont {G.}~\bibnamefont {Madariaga}}, \bibinfo
  {author} {\bibfnamefont {J.~M.}\ \bibnamefont {P{\'e}rez-Mato}}, \ and\
  \bibinfo {author} {\bibfnamefont {M.~I.}\ \bibnamefont {Aroyo}},\ }\href@noop
  {} {\bibfield  {journal} {\bibinfo  {journal} {J. Appl. Cryst.}\ }\textbf
  {\bibinfo {volume} {33}},\ \bibinfo {pages} {1190} (\bibinfo {year}
  {2000})}\BibitemShut {NoStop}%
\bibitem [{\citenamefont {Aroyo}\ \emph
  {et~al.}(2006{\natexlab{a}})\citenamefont {Aroyo}, \citenamefont
  {P{\'e}rez-Mato}, \citenamefont {Capillas}, \citenamefont {Kroumova},
  \citenamefont {Ivantchev}, \citenamefont {Madariaga}, \citenamefont {Kirov},\
  and\ \citenamefont {Wondratschek}}]{2006aAroyo}%
  \BibitemOpen
  \bibfield  {author} {\bibinfo {author} {\bibfnamefont {M.}~\bibnamefont
  {Aroyo}}, \bibinfo {author} {\bibfnamefont {J.~M.}\ \bibnamefont
  {P{\'e}rez-Mato}}, \bibinfo {author} {\bibfnamefont {C.}~\bibnamefont
  {Capillas}}, \bibinfo {author} {\bibfnamefont {E.}~\bibnamefont {Kroumova}},
  \bibinfo {author} {\bibfnamefont {S.}~\bibnamefont {Ivantchev}}, \bibinfo
  {author} {\bibfnamefont {G.}~\bibnamefont {Madariaga}}, \bibinfo {author}
  {\bibfnamefont {A.}~\bibnamefont {Kirov}}, \ and\ \bibinfo {author}
  {\bibfnamefont {H.}~\bibnamefont {Wondratschek}},\ }\href@noop {} {\bibfield
  {journal} {\bibinfo  {journal} {Z. Krist.}\ }\textbf {\bibinfo {volume}
  {221}},\ \bibinfo {pages} {15} (\bibinfo {year}
  {2006}{\natexlab{a}})}\BibitemShut {NoStop}%
\bibitem [{\citenamefont {Aroyo}\ \emph
  {et~al.}(2006{\natexlab{b}})\citenamefont {Aroyo}, \citenamefont {Capillas},
  \citenamefont {P{\'e}rez-Mato},\ and\ \citenamefont
  {Wondratschek}}]{2006bAroyo}%
  \BibitemOpen
  \bibfield  {author} {\bibinfo {author} {\bibfnamefont {M.}~\bibnamefont
  {Aroyo}}, \bibinfo {author} {\bibfnamefont {C.}~\bibnamefont {Capillas}},
  \bibinfo {author} {\bibfnamefont {J.~M.}\ \bibnamefont {P{\'e}rez-Mato}}, \
  and\ \bibinfo {author} {\bibfnamefont {H.}~\bibnamefont {Wondratschek}},\
  }\href@noop {} {\bibfield  {journal} {\bibinfo  {journal} {Acta Cryst. A}\
  }\textbf {\bibinfo {volume} {62}},\ \bibinfo {pages} {115} (\bibinfo {year}
  {2006}{\natexlab{b}})}\BibitemShut {NoStop}%
\bibitem [{\citenamefont {Aroyo}\ \emph {et~al.}(2011)\citenamefont {Aroyo},
  \citenamefont {P{\'e}rez-Mato}, \citenamefont {Orobengoa}, \citenamefont
  {Tasci}, \citenamefont {de~la Flor},\ and\ \citenamefont
  {Kirov}}]{2011Aroyo}%
  \BibitemOpen
  \bibfield  {author} {\bibinfo {author} {\bibfnamefont {M.~I.}\ \bibnamefont
  {Aroyo}}, \bibinfo {author} {\bibfnamefont {J.~M.}\ \bibnamefont
  {P{\'e}rez-Mato}}, \bibinfo {author} {\bibfnamefont {D.}~\bibnamefont
  {Orobengoa}}, \bibinfo {author} {\bibfnamefont {E.}~\bibnamefont {Tasci}},
  \bibinfo {author} {\bibfnamefont {G.}~\bibnamefont {de~la Flor}}, \ and\
  \bibinfo {author} {\bibfnamefont {A.}~\bibnamefont {Kirov}},\ }\href@noop {}
  {\bibfield  {journal} {\bibinfo  {journal} {Bulg. Chem. Commun.}\ }\textbf
  {\bibinfo {volume} {43(2)}},\ \bibinfo {pages} {183} (\bibinfo {year}
  {2011})}\BibitemShut {NoStop}%
\bibitem [{\citenamefont {Hushur}\ \emph {et~al.}(2004)\citenamefont {Hushur},
  \citenamefont {Shabbir}, \citenamefont {Ko},\ and\ \citenamefont
  {Kojima}}]{2004Hushur}%
  \BibitemOpen
  \bibfield  {author} {\bibinfo {author} {\bibfnamefont {A.}~\bibnamefont
  {Hushur}}, \bibinfo {author} {\bibfnamefont {G.}~\bibnamefont {Shabbir}},
  \bibinfo {author} {\bibfnamefont {J.-H.}\ \bibnamefont {Ko}}, \ and\ \bibinfo
  {author} {\bibfnamefont {S.}~\bibnamefont {Kojima}},\ }\href
  {http://stacks.iop.org/0022-3727/37/i=7/a=028} {\bibfield  {journal}
  {\bibinfo  {journal} {J. Phys. D}\ }\textbf {\bibinfo {volume} {37}},\
  \bibinfo {pages} {1127} (\bibinfo {year} {2004})}\BibitemShut {NoStop}%
\bibitem [{\citenamefont {Ederer}\ and\ \citenamefont
  {Spaldin}(2006)}]{ederer}%
  \BibitemOpen
  \bibfield  {author} {\bibinfo {author} {\bibfnamefont {C.}~\bibnamefont
  {Ederer}}\ and\ \bibinfo {author} {\bibfnamefont {N.~A.}\ \bibnamefont
  {Spaldin}},\ }\href@noop {} {\bibfield  {journal} {\bibinfo  {journal} {Phys.
  Rev. B}\ }\textbf {\bibinfo {volume} {74}},\ \bibinfo {pages} {024102}
  (\bibinfo {year} {2006})}\BibitemShut {NoStop}%
\bibitem [{\citenamefont {N{\'u\~n}ez~Valdez}\ \emph
  {et~al.}(2016)\citenamefont {N{\'u\~n}ez~Valdez}, \citenamefont {Spanke},\
  and\ \citenamefont {Spaldin}}]{2016MNV}%
  \BibitemOpen
  \bibfield  {author} {\bibinfo {author} {\bibfnamefont {M.}~\bibnamefont
  {N{\'u\~n}ez~Valdez}}, \bibinfo {author} {\bibfnamefont {H.~T.}\ \bibnamefont
  {Spanke}}, \ and\ \bibinfo {author} {\bibfnamefont {N.~A.}\ \bibnamefont
  {Spaldin}},\ }\href@noop {} {\bibfield  {journal} {\bibinfo  {journal} {Phys.
  Rev. B}\ }\textbf {\bibinfo {volume} {93}},\ \bibinfo {pages} {64112}
  (\bibinfo {year} {2016})}\BibitemShut {NoStop}%
\bibitem [{\citenamefont {Bruyer}\ and\ \citenamefont {Sayede}(2010)}]{bruyer}%
  \BibitemOpen
  \bibfield  {author} {\bibinfo {author} {\bibfnamefont {E.}~\bibnamefont
  {Bruyer}}\ and\ \bibinfo {author} {\bibfnamefont {A.}~\bibnamefont
  {Sayede}},\ }\href@noop {} {\bibfield  {journal} {\bibinfo  {journal} {J.
  Appl. Phys.}\ }\textbf {\bibinfo {volume} {108}},\ \bibinfo {pages} {053705}
  (\bibinfo {year} {2010})}\BibitemShut {NoStop}%
\bibitem [{\citenamefont {Hohenberg}\ and\ \citenamefont {Kohn}(1964)}]{dft1}%
  \BibitemOpen
  \bibfield  {author} {\bibinfo {author} {\bibfnamefont {P.}~\bibnamefont
  {Hohenberg}}\ and\ \bibinfo {author} {\bibfnamefont {W.}~\bibnamefont
  {Kohn}},\ }\href@noop {} {\bibfield  {journal} {\bibinfo  {journal} {Phys.\
  Rev. B}\ }\textbf {\bibinfo {volume} {136}},\ \bibinfo {pages} {864}
  (\bibinfo {year} {1964})}\BibitemShut {NoStop}%
\bibitem [{\citenamefont {Kohn}\ and\ \citenamefont {Sham}(1964)}]{dft2}%
  \BibitemOpen
  \bibfield  {author} {\bibinfo {author} {\bibfnamefont {W.}~\bibnamefont
  {Kohn}}\ and\ \bibinfo {author} {\bibfnamefont {L.~J.}\ \bibnamefont
  {Sham}},\ }\href@noop {} {\bibfield  {journal} {\bibinfo  {journal} {Phys.\
  Rev. A}\ }\textbf {\bibinfo {volume} {140}},\ \bibinfo {pages} {1133}
  (\bibinfo {year} {1964})}\BibitemShut {NoStop}%
\bibitem [{\citenamefont {Kresse}\ and\ \citenamefont
  {Hafner}(1993)}]{kresse93}%
  \BibitemOpen
  \bibfield  {author} {\bibinfo {author} {\bibfnamefont {G.}~\bibnamefont
  {Kresse}}\ and\ \bibinfo {author} {\bibfnamefont {J.}~\bibnamefont
  {Hafner}},\ }\href@noop {} {\bibfield  {journal} {\bibinfo  {journal} {Phys.\
  Rev.\ B}\ }\textbf {\bibinfo {volume} {47}},\ \bibinfo {pages} {558}
  (\bibinfo {year} {1993})}\BibitemShut {NoStop}%
\bibitem [{\citenamefont {Kresse}\ and\ \citenamefont
  {Hafner}(1994)}]{kresse94}%
  \BibitemOpen
  \bibfield  {author} {\bibinfo {author} {\bibfnamefont {G.}~\bibnamefont
  {Kresse}}\ and\ \bibinfo {author} {\bibfnamefont {J.}~\bibnamefont
  {Hafner}},\ }\href@noop {} {\bibfield  {journal} {\bibinfo  {journal} {Phys.\
  Rev.\ B}\ }\textbf {\bibinfo {volume} {49}},\ \bibinfo {pages} {14251}
  (\bibinfo {year} {1994})}\BibitemShut {NoStop}%
\bibitem [{\citenamefont {Kresse}\ and\ \citenamefont
  {Furthm{\"{u}}ller}(1996{\natexlab{a}})}]{kresse96a}%
  \BibitemOpen
  \bibfield  {author} {\bibinfo {author} {\bibfnamefont {G.}~\bibnamefont
  {Kresse}}\ and\ \bibinfo {author} {\bibfnamefont {J.}~\bibnamefont
  {Furthm{\"{u}}ller}},\ }\href@noop {} {\bibfield  {journal} {\bibinfo
  {journal} {Comp. Mater. Sci.}\ }\textbf {\bibinfo {volume} {6}},\ \bibinfo
  {pages} {15} (\bibinfo {year} {1996}{\natexlab{a}})}\BibitemShut {NoStop}%
\bibitem [{\citenamefont {Kresse}\ and\ \citenamefont
  {Furthm{\"{u}}ller}(1996{\natexlab{b}})}]{kresse96b}%
  \BibitemOpen
  \bibfield  {author} {\bibinfo {author} {\bibfnamefont {G.}~\bibnamefont
  {Kresse}}\ and\ \bibinfo {author} {\bibfnamefont {J.}~\bibnamefont
  {Furthm{\"{u}}ller}},\ }\href@noop {} {\bibfield  {journal} {\bibinfo
  {journal} {Phys.\ Rev.\ B}\ }\textbf {\bibinfo {volume} {54}},\ \bibinfo
  {pages} {11169} (\bibinfo {year} {1996}{\natexlab{b}})}\BibitemShut {NoStop}%
\bibitem [{\citenamefont {Perdew}\ and\ \citenamefont {Wang}(1992)}]{perdew1}%
  \BibitemOpen
  \bibfield  {author} {\bibinfo {author} {\bibfnamefont {J.~P.}\ \bibnamefont
  {Perdew}}\ and\ \bibinfo {author} {\bibfnamefont {Y.}~\bibnamefont {Wang}},\
  }\href@noop {} {\bibfield  {journal} {\bibinfo  {journal} {Phys.\ Rev.\ B}\
  }\textbf {\bibinfo {volume} {45}},\ \bibinfo {pages} {13244} (\bibinfo {year}
  {1992})}\BibitemShut {NoStop}%
\bibitem [{\citenamefont {Perdew}\ \emph {et~al.}(1996)\citenamefont {Perdew},
  \citenamefont {Burke},\ and\ \citenamefont {Ernzerhof}}]{perdew2}%
  \BibitemOpen
  \bibfield  {author} {\bibinfo {author} {\bibfnamefont {J.~P.}\ \bibnamefont
  {Perdew}}, \bibinfo {author} {\bibfnamefont {K.}~\bibnamefont {Burke}}, \
  and\ \bibinfo {author} {\bibfnamefont {M.}~\bibnamefont {Ernzerhof}},\
  }\href@noop {} {\bibfield  {journal} {\bibinfo  {journal} {Phys.\ Rev.\
  Lett.}\ }\textbf {\bibinfo {volume} {77}},\ \bibinfo {pages} {3865} (\bibinfo
  {year} {1996})}\BibitemShut {NoStop}%
\bibitem [{\citenamefont {Perdew}\ \emph {et~al.}(2008)\citenamefont {Perdew},
  \citenamefont {Ruzsinszky}, \citenamefont {Csonka}, \citenamefont {Vydrov},
  \citenamefont {Scuseria}, \citenamefont {Constantin}, \citenamefont {Zhou},\
  and\ \citenamefont {Burke}}]{PBEsol}%
  \BibitemOpen
  \bibfield  {author} {\bibinfo {author} {\bibfnamefont {J.}~\bibnamefont
  {Perdew}}, \bibinfo {author} {\bibfnamefont {A.}~\bibnamefont {Ruzsinszky}},
  \bibinfo {author} {\bibfnamefont {G.}~\bibnamefont {Csonka}}, \bibinfo
  {author} {\bibfnamefont {O.}~\bibnamefont {Vydrov}}, \bibinfo {author}
  {\bibfnamefont {G.}~\bibnamefont {Scuseria}}, \bibinfo {author}
  {\bibfnamefont {L.}~\bibnamefont {Constantin}}, \bibinfo {author}
  {\bibfnamefont {X.}~\bibnamefont {Zhou}}, \ and\ \bibinfo {author}
  {\bibfnamefont {K.}~\bibnamefont {Burke}},\ }\href@noop {} {\bibfield
  {journal} {\bibinfo  {journal} {Phys. Rev. Lett.}\ }\textbf {\bibinfo
  {volume} {100}},\ \bibinfo {pages} {136406} (\bibinfo {year}
  {2008})}\BibitemShut {NoStop}%
\bibitem [{\citenamefont {Bl{\"{o}}chl}(1994)}]{blochl}%
  \BibitemOpen
  \bibfield  {author} {\bibinfo {author} {\bibfnamefont {P.~E.}\ \bibnamefont
  {Bl{\"{o}}chl}},\ }\href@noop {} {\bibfield  {journal} {\bibinfo  {journal}
  {Phys. Rev. B}\ }\textbf {\bibinfo {volume} {50}},\ \bibinfo {pages} {17953}
  (\bibinfo {year} {1994})}\BibitemShut {NoStop}%
\bibitem [{\citenamefont {Kresse}\ and\ \citenamefont
  {Joubert}(1999)}]{kresse99}%
  \BibitemOpen
  \bibfield  {author} {\bibinfo {author} {\bibfnamefont {G.}~\bibnamefont
  {Kresse}}\ and\ \bibinfo {author} {\bibfnamefont {D.}~\bibnamefont
  {Joubert}},\ }\href@noop {} {\bibfield  {journal} {\bibinfo  {journal}
  {Phys.\ Rev.\ B}\ }\textbf {\bibinfo {volume} {59}},\ \bibinfo {pages} {1758}
  (\bibinfo {year} {1999})}\BibitemShut {NoStop}%
\bibitem [{\citenamefont {Monkhorst}\ and\ \citenamefont {Pack}(1976)}]{kmesh}%
  \BibitemOpen
  \bibfield  {author} {\bibinfo {author} {\bibfnamefont {H.~J.}\ \bibnamefont
  {Monkhorst}}\ and\ \bibinfo {author} {\bibfnamefont {J.~D.}\ \bibnamefont
  {Pack}},\ }\href@noop {} {\bibfield  {journal} {\bibinfo  {journal} {Phys.
  Rev. B,}\ }\textbf {\bibinfo {volume} {13}},\ \bibinfo {pages} {5188}
  (\bibinfo {year} {1976})}\BibitemShut {NoStop}%
\bibitem [{\citenamefont {King-Smith}\ and\ \citenamefont
  {Vanderbilt}(1993)}]{berry1}%
  \BibitemOpen
  \bibfield  {author} {\bibinfo {author} {\bibfnamefont {R.~D.}\ \bibnamefont
  {King-Smith}}\ and\ \bibinfo {author} {\bibfnamefont {D.}~\bibnamefont
  {Vanderbilt}},\ }\href@noop {} {\bibfield  {journal} {\bibinfo  {journal}
  {Phys. Rev. B}\ }\textbf {\bibinfo {volume} {47}},\ \bibinfo {pages} {R1651}
  (\bibinfo {year} {1993})}\BibitemShut {NoStop}%
\bibitem [{\citenamefont {Vanderbilt}\ and\ \citenamefont
  {King-Smith}(1994)}]{berry2}%
  \BibitemOpen
  \bibfield  {author} {\bibinfo {author} {\bibfnamefont {D.}~\bibnamefont
  {Vanderbilt}}\ and\ \bibinfo {author} {\bibfnamefont {R.~D.}\ \bibnamefont
  {King-Smith}},\ }\href@noop {} {\bibfield  {journal} {\bibinfo  {journal}
  {Phys. Rev. B}\ }\textbf {\bibinfo {volume} {48}},\ \bibinfo {pages} {4442}
  (\bibinfo {year} {1994})}\BibitemShut {NoStop}%
\bibitem [{\citenamefont {Resta}(1994)}]{berry3}%
  \BibitemOpen
  \bibfield  {author} {\bibinfo {author} {\bibfnamefont {R.}~\bibnamefont
  {Resta}},\ }\href@noop {} {\bibfield  {journal} {\bibinfo  {journal} {Rev.
  Mod. Phys.}\ }\textbf {\bibinfo {volume} {66}},\ \bibinfo {pages} {899}
  (\bibinfo {year} {1994})}\BibitemShut {NoStop}%
\bibitem [{\citenamefont {Spaldin}(2012)}]{spaldin}%
  \BibitemOpen
  \bibfield  {author} {\bibinfo {author} {\bibfnamefont {N.~A.}\ \bibnamefont
  {Spaldin}},\ }\href@noop {} {\bibfield  {journal} {\bibinfo  {journal} {J.
  Sol. Stat. Chem.}\ }\textbf {\bibinfo {volume} {195}},\ \bibinfo {pages} {2}
  (\bibinfo {year} {2012})}\BibitemShut {NoStop}%
\bibitem [{\citenamefont {Zhang}\ \emph {et~al.}(2017)\citenamefont {Zhang},
  \citenamefont {Sun}, \citenamefont {Perdew},\ and\ \citenamefont
  {Wu}}]{2017Zhang}%
  \BibitemOpen
  \bibfield  {author} {\bibinfo {author} {\bibfnamefont {Y.}~\bibnamefont
  {Zhang}}, \bibinfo {author} {\bibfnamefont {J.}~\bibnamefont {Sun}}, \bibinfo
  {author} {\bibfnamefont {J.~P.}\ \bibnamefont {Perdew}}, \ and\ \bibinfo
  {author} {\bibfnamefont {X.}~\bibnamefont {Wu}},\ }\href@noop {} {\bibfield
  {journal} {\bibinfo  {journal} {Phys. Rev. B}\ }\textbf {\bibinfo {volume}
  {96}},\ \bibinfo {pages} {035143} (\bibinfo {year} {2017})}\BibitemShut
  {NoStop}%
\bibitem [{\citenamefont {Hirosaki}\ \emph {et~al.}(2003)\citenamefont
  {Hirosaki}, \citenamefont {Ogata},\ and\ \citenamefont
  {Kocer}}]{2003Hirosaki}%
  \BibitemOpen
  \bibfield  {author} {\bibinfo {author} {\bibfnamefont {N.}~\bibnamefont
  {Hirosaki}}, \bibinfo {author} {\bibfnamefont {S.}~\bibnamefont {Ogata}}, \
  and\ \bibinfo {author} {\bibfnamefont {C.}~\bibnamefont {Kocer}},\
  }\href@noop {} {\bibfield  {journal} {\bibinfo  {journal} {J. Alloys Compd.}\
  }\textbf {\bibinfo {volume} {351}},\ \bibinfo {pages} {31} (\bibinfo {year}
  {2003})}\BibitemShut {NoStop}%
\bibitem [{\citenamefont {Wu}\ \emph {et~al.}(2007)\citenamefont {Wu},
  \citenamefont {Zinkevich}, \citenamefont {Aldinger}, \citenamefont {Wen},\
  and\ \citenamefont {Chen}}]{2007Wu}%
  \BibitemOpen
  \bibfield  {author} {\bibinfo {author} {\bibfnamefont {B.}~\bibnamefont
  {Wu}}, \bibinfo {author} {\bibfnamefont {M.}~\bibnamefont {Zinkevich}},
  \bibinfo {author} {\bibfnamefont {F.}~\bibnamefont {Aldinger}}, \bibinfo
  {author} {\bibfnamefont {D.}~\bibnamefont {Wen}}, \ and\ \bibinfo {author}
  {\bibfnamefont {L.}~\bibnamefont {Chen}},\ }\href@noop {} {\bibfield
  {journal} {\bibinfo  {journal} {J. Solid State Chem.}\ }\textbf {\bibinfo
  {volume} {180}},\ \bibinfo {pages} {3280} (\bibinfo {year}
  {2007})}\BibitemShut {NoStop}%
\bibitem [{\citenamefont {Orobengoa}\ \emph {et~al.}(2009)\citenamefont
  {Orobengoa}, \citenamefont {Capillas}, \citenamefont {Aroyo},\ and\
  \citenamefont {Perez-Mato}}]{orobengoa}%
  \BibitemOpen
  \bibfield  {author} {\bibinfo {author} {\bibfnamefont {D.}~\bibnamefont
  {Orobengoa}}, \bibinfo {author} {\bibfnamefont {C.}~\bibnamefont {Capillas}},
  \bibinfo {author} {\bibfnamefont {M.}~\bibnamefont {Aroyo}}, \ and\ \bibinfo
  {author} {\bibfnamefont {J.}~\bibnamefont {Perez-Mato}},\ }\href@noop {}
  {\bibfield  {journal} {\bibinfo  {journal} {J. Appl. Cryst.}\ }\textbf
  {\bibinfo {volume} {A42}},\ \bibinfo {pages} {820} (\bibinfo {year}
  {2009})}\BibitemShut {NoStop}%
\bibitem [{\citenamefont {Perez-Mato}\ \emph {et~al.}(2010)\citenamefont
  {Perez-Mato}, \citenamefont {Orobengoa},\ and\ \citenamefont
  {Aroyo}}]{perez}%
  \BibitemOpen
  \bibfield  {author} {\bibinfo {author} {\bibfnamefont {J.}~\bibnamefont
  {Perez-Mato}}, \bibinfo {author} {\bibfnamefont {D.}~\bibnamefont
  {Orobengoa}}, \ and\ \bibinfo {author} {\bibfnamefont {M.}~\bibnamefont
  {Aroyo}},\ }\href@noop {} {\bibfield  {journal} {\bibinfo  {journal} {Acta
  Cryst. A}\ }\textbf {\bibinfo {volume} {66}},\ \bibinfo {pages} {558}
  (\bibinfo {year} {2010})}\BibitemShut {NoStop}%
\bibitem [{\citenamefont {Van~Aken}\ \emph {et~al.}(2004)\citenamefont
  {Van~Aken}, \citenamefont {Palstra}, \citenamefont {Filippetti},\ and\
  \citenamefont {Spaldin}}]{2004vanAken}%
  \BibitemOpen
  \bibfield  {author} {\bibinfo {author} {\bibfnamefont {B.}~\bibnamefont
  {Van~Aken}}, \bibinfo {author} {\bibfnamefont {T.}~\bibnamefont {Palstra}},
  \bibinfo {author} {\bibfnamefont {A.}~\bibnamefont {Filippetti}}, \ and\
  \bibinfo {author} {\bibfnamefont {N.~A.}\ \bibnamefont {Spaldin}},\
  }\href@noop {} {\bibfield  {journal} {\bibinfo  {journal} {Nat. Mater.}\
  }\textbf {\bibinfo {volume} {3}},\ \bibinfo {pages} {164} (\bibinfo {year}
  {2004})}\BibitemShut {NoStop}%
\end{thebibliography}%
\appendix
\section{Born effective charge tensors}
\begin{figure*}
\includegraphics[width=0.4\textwidth]{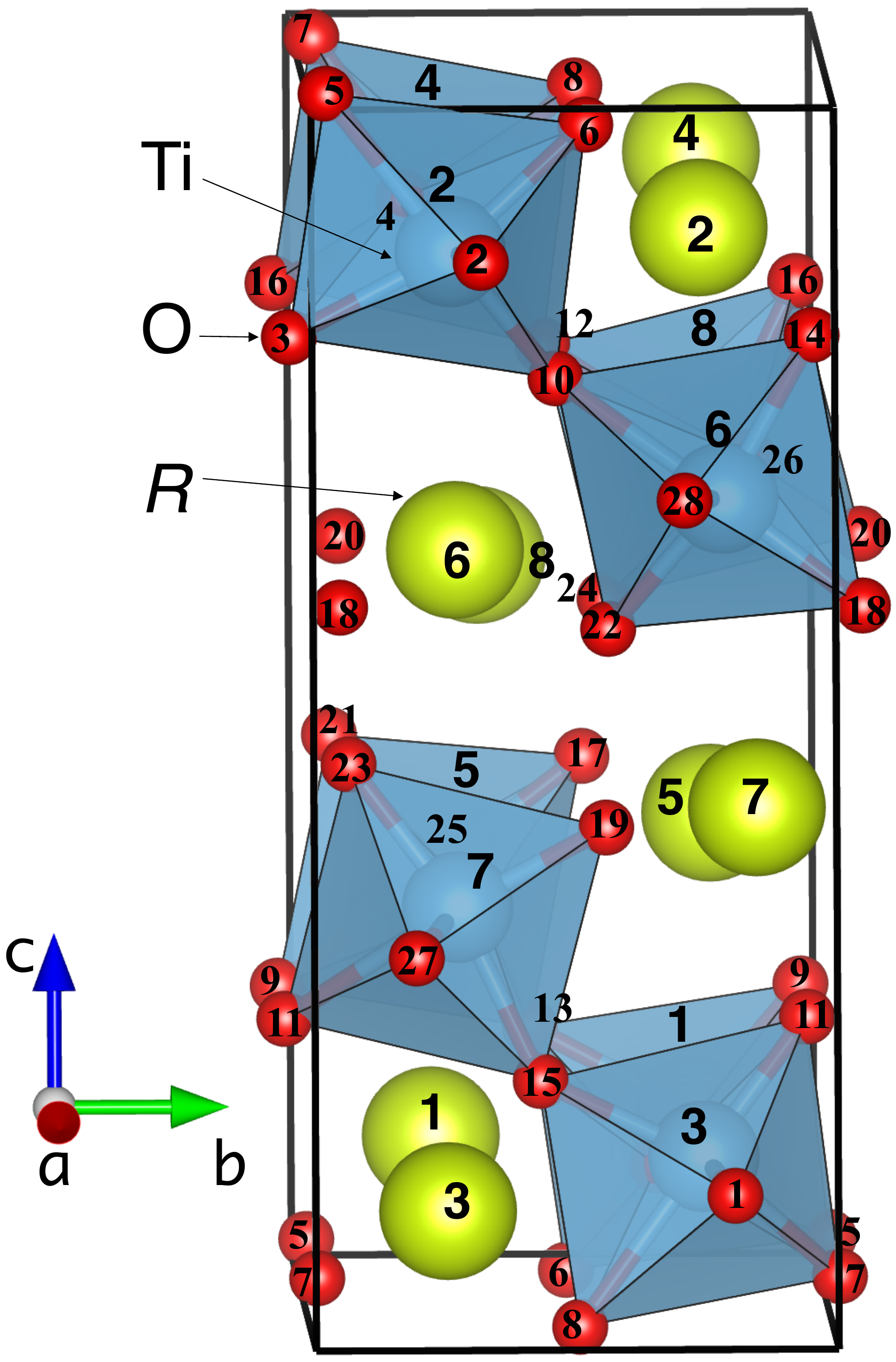}
\caption{\label{figap01} The ferroelectric $P2_1$ structure indicating the numbering scheme for the atoms. The atomic numbering is the same in the centrosymmetric $P2_1/m$ phase.} 
\end{figure*}

Born effective charge tensors calculated in this work using the PBEsol functional (given in units of elementary charge $e$) for the polar $P2_1$ and the centrosymmetric reference $P2_1/m$ structures of $R_2$Ti$_2$O$_7$ with $R=$ La, Ce, Pr, and Nd. The numbering of the atoms in Tables IV-VII is indicated in Fig. \ref{figap01}.
\newpage
\begin{table*}
{
\caption{\label{BCval}La$_2$Ti$_2$O$_7$ Born effective charge tensors in units of $e$.}
\begin{tabular}{lclclc}
\hline
\hline
Atom & ${P2_1}$ & Atom & ${P2_1}$ &Atom &  $P2_1/m$ \\
\hline
La(1,2) & 
$
 \begin{pmatrix*}[r]
4.28	&	\pm0.02	&	0.01	\\
\pm0.04	&	4.56	&	\pm0.30	\\
-0.03	&	\pm0.20	&	3.31	
 \end{pmatrix*}
$
&
O(1,2) &
$\begin{pmatrix*}[r]					
-5.51	&	0.00	&	0.02	\\
0.00	&	-2.38	&	\pm0.02	\\
0.00	&	\pm0.05	&	-1.34	
\end{pmatrix*} $ &
La(1,\dots,4) &
$\begin{pmatrix*}[r]					
4.39	&	0.00	&	0.00	\\
0.00	&	4.21	&	0.00	\\
0.00	&	0.00	&	3.66	
\end{pmatrix*} $ 
\\ \\
La(3,4) &
 $ \begin{pmatrix*}[r]					
4.38	&	\pm0.02	&	0.01	\\
\pm0.05	&	4.19	&	\pm0.59	\\
0.02	&	\pm0.09	&	3.83	
\end{pmatrix*} $
 &
 O(3,4) &
$\begin{pmatrix*}[r]
-4.70	&	\pm0.02	&	-0.02	\\
0.00	&	-2.54	&	\pm0.07	\\
0.00	&	\pm0.15	&	-1.82	
\end{pmatrix*} $ &
La(5,\dots,8) &
$\begin{pmatrix*}[r]	
4.61	&	0.00	&	0.00	\\
0.00	&	4.67	&	0.00	\\
0.00	&	0.00	&	4.20	
\end{pmatrix*}$
\\ \\
La(5,6) &
$\begin{pmatrix*}[r]					
4.54	&	\pm0.07	&	-0.06	\\
\pm0.19	&	4.61	&	\pm0.12	\\
-0.09	&	\pm0.59	&	4.45	
\end{pmatrix*}$ &
O(5,6) &
$\begin{pmatrix*}[r]
-2.46	&	0.00	&	-0.09	\\
\pm0.09	&	-3.40	&	\pm1.10	\\
0.03	&	\pm0.99	&	-2.96	
\end{pmatrix*} $ &
Ti(1,\dots,4) & 
$\begin{pmatrix*}[r]					
6.67	&	0.00	&	0.00	\\
0.00	&	7.91	&	0.00	\\
0.00	&	0.00	&	5.44	
\end{pmatrix*}$
\\ \\
La(7,8) &
$\begin{pmatrix*}[r]					
4.47	&	\pm0.11	&	0.04	\\
\pm0.16	&	4.72	&	\pm0.34	\\
0.11	&	\pm0.50	&	4.39		
\end{pmatrix*}$ &
O(7,8) &
$\begin{pmatrix*}[r]
-2.49	&	\pm0.01	&	0.09	\\
\pm0.10	&	-3.34	&	\pm1.03	\\
-0.02	&	\pm0.91	&	-2.88	
\end{pmatrix*} $ &
Ti(5,\dots,8) &
$\begin{pmatrix*}[r]					
7.42	&	0.00	&	0.00	\\
0.00	&	5.80	&	0.00	\\
0.00	&	0.00	&	7.27	
\end{pmatrix*} $ 
\\ \\
Ti(1,2) &
$ \begin{pmatrix*}[r]					
6.44	&	\pm0.07	&	0.21	\\
\pm0.33	&	6.63	&	\pm0.09	\\
0.20	&	\pm0.34	&	5.00	
\end{pmatrix*} $ & 
O(9,10) &
$\begin{pmatrix*}[r]
-1.83	&	\pm0.31	&	-0.31	\\
\pm0.27	&	-3.20	&	\pm1.46	\\
-0.32	&	\pm1.37	&	-3.22	
\end{pmatrix*} $ &
O(1,\dots,4) &
$\begin{pmatrix*}[r]					
-5.42	&	0.00	&	0.00	\\
0.00	&	-2.41	&	0.00	\\
0.00	&	0.00	&	-1.68	
\end{pmatrix*} $ 
\\ \\
Ti(3,4) &
$\begin{pmatrix*}[r]					
6.37	&	\pm0.07	&	-0.20	\\
\pm0.26	&	6.42	&	\pm0.05	\\
-0.20	&	\pm0.28	&	4.78	
\end{pmatrix*} $ & 
O(11,12) &
$\begin{pmatrix*}[r]
-1.81	&	\pm0.28	&	0.36	\\
\pm0.25	&	-3.10	&	\pm1.36	\\
0.40	&	\pm1.22	&	-3.21	
\end{pmatrix*} $ &
O(5,\dots,8) &
$\begin{pmatrix*}[r]					
-2.47	&	0.00	&	0.00	\\
0.00	&	-3.83	&	\pm1.34	\\
0.00	&	\pm1.08	&	-3.20	
\end{pmatrix*} $ 
\\ \\
Ti(5,6) &
$\begin{pmatrix*}[r]					
6.44	&	\pm0.88	&	0.13	\\
\pm0.06	&	5.50	&	\pm0.52	\\
-0.17	&	\pm0.28	&	6.07	
\end{pmatrix*} $ & 
O(13,14) &
$\begin{pmatrix*}[r]
-2.39	&	\pm0.23	&	-0.22	\\
\pm0.17	&	-3.88	&	\pm1.45	\\
-0.12	&	\pm1.37	&	-3.23	
\end{pmatrix*} $ &
O(9,\dots,16) &
$\begin{pmatrix*}[r]					
-2.10	&	0.00	&	0.00	\\
0.00	&	-4.13	&	\pm1.88	\\
0.00	&	\pm1.68	&	-3.40	
\end{pmatrix*} $ 
\\ \\
Ti(7,8) &
$\begin{pmatrix*}[r]					
6.28	&\pm0.84	&	-0.03	\\
\pm0.13	&	5.76	&	\pm0.51	\\
0.24	&	\pm0.05	&	5.49	
\end{pmatrix*} $ & 
O(15,16) &
$\begin{pmatrix*}[r]
-2.47	&	\pm0.26	&	0.22	\\
\pm0.18	&	-3.82	&	\pm1.18	\\
0.13	&	\pm1.23	&	-2.97	
\end{pmatrix*} $ &
O(17,\dots,20) &
$ \begin{pmatrix*}[r]					
-5.95	&	0.00	&	0.00 \\
0.00	&	-1.59	&	0.00 \\
0.00	&	0.00	&	-2.18
\end{pmatrix*} $ 
\\ \\
 & & O(17,18) &
$\begin{pmatrix*}[r]
-2.16	&	\pm0.19	&	0.17	\\
\pm0.20	&	-3.27	&	\pm0.88	\\
0.22	&	\pm0.86	&	-2.91	
\end{pmatrix*} $ & 
O(21,\dots,28) &
$ \begin{pmatrix*}[r]					
-2.53	&	0.00	&	0.00\\
0.00	&	-3.26	&	\pm0.76\\
0.00	&	\pm0.94	&	-3.36
\end{pmatrix*} $ 
\\ \\
 & & O(19,20) &
$\begin{pmatrix*}[r]
-2.43	&	\pm0.26	&	-0.23	\\
\pm0.30	&	-3.76	&	\pm1.12	\\
-0.33	&	\pm1.17	&	-1.95	
\end{pmatrix*} $ &
 \\ \\
 & & O(21,22) &
$\begin{pmatrix*}[r]
-2.38	&	\pm0.10	&	0.08	\\
\pm0.21	&	-3.08	&	\pm0.68	\\
0.10	&	\pm0.48	&	-3.28	
\end{pmatrix*} $ &
& \\ \\
 & & O(23,24) &
$\begin{pmatrix*}[r]
-2.25	&	\pm0.02	&	-0.09	\\
\pm0.10	&	-2.95	&	\pm0.81	\\
-0.12	&	\pm0.50	&	-3.55	
\end{pmatrix*} $ &
& \\ \\
 & & O(25,26) &
$\begin{pmatrix*}[r]
-5.14	&	\pm0.03	&	-0.08	\\
\pm0.08	&	-1.88	&	\pm0.19	\\
-0.05	&	\pm0.32	&	-1.99		
\end{pmatrix*} $ &
& \\ \\
 & & O(27,28) &
$\begin{pmatrix*}[r]
-5.17	&	\pm0.05	&	0.02	\\
\pm0.04	&	-1.80	&	\pm0.40	\\
0.00	&	\pm0.36	&	-2.01	
\end{pmatrix*} $ &
&  \\
\hline 
\hline
\end{tabular}
}
\end{table*}
\newpage
\begin{table*}
{
\caption{\label{BCval}Ce$_2$Ti$_2$O$_7$ Born effective charge tensors in units of $e$.}
\begin{tabular}{lclclc}
\hline
\hline
Atom & ${P2_1}$ & Atom & ${P2_1}$ &Atom &  $P2_1/m$ \\
\hline
Ce(1,2) & 
$
 \begin{pmatrix*}[r]
 4.05	&	\pm0.03	&	0.02	\\
\pm0.05	&	4.50	&	\pm0.30	\\
-0.04	&	\pm0.19	&	3.23	
 \end{pmatrix*}
$
&
O(1,2) &
$\begin{pmatrix*}[r]					
-5.59	&	0.00	&	0.02	\\
0.00	&	-2.31	&	\pm0.04	\\
0.00	&	\pm0.08	&	-1.33	
\end{pmatrix*} $ &
Ce(1,\dots,4) &
$\begin{pmatrix*}[r]					
4.38	&	0.00	&	0.00	\\
0.00	&	4.25	&	0.00	\\
0.00	&	0.00	&	3.63	\\
\end{pmatrix*} $ 
\\ \\
Ce(3,4) &
 $ \begin{pmatrix*}[r]					
4.22	&	\pm0.03	&	0.02	\\
\pm0.05	&	4.11	&	\pm0.58	\\
0.03	&	\pm0.05	&	3.76	
\end{pmatrix*} $
 &
 O(3,4) &
$\begin{pmatrix*}[r]
-4.71	&	\pm0.02	&	-0.02	\\
0.00	&	-2.45	&	\pm0.07	\\
0.01	&	\pm0.15	&	-1.78	
\end{pmatrix*} $ &
Ce(5,\dots,8) &
$\begin{pmatrix*}[r]	
4.47	&	0.00	&	0.00	\\
0.00	&	4.66	&	0.00	\\
0.00	&	0.00	&	4.23	\\
\end{pmatrix*}$
\\ \\
Ce(5,6) &
$\begin{pmatrix*}[r]					
4.31	&	\pm0.06	&	-0.08	\\
\pm0.19	&	4.53	&	\pm0.14	\\
-0.08	&	\pm0.62	&	4.39	
\end{pmatrix*}$ &
O(5,6) &
$\begin{pmatrix*}[r]
-2.40	&	\pm0.02	&	-0.06	\\
\pm0.08	&	-3.36	&	\pm1.08	\\
0.05	&	\pm0.98	&	-2.91	
\end{pmatrix*} $ &
Ti(1,\dots,4) & 
$\begin{pmatrix*}[r]					
6.69	&	0.00	&	0.00	\\
0.00	&	7.98	&	0.00	\\
0.00	&	0.00	&	5.46	
\end{pmatrix*}$
\\ \\
Ce(7,8) &
$\begin{pmatrix*}[r]					
4.26	&	\pm0.11	&	0.06	\\
\pm0.16	&	4.63	&	\pm0.29	\\
0.14	&	\pm0.50	&	4.40	
\end{pmatrix*}$ &
O(7,8) &
$\begin{pmatrix*}[r]
-2.44	&	\pm0.01	&	0.04	\\
\pm0.08	&	-3.30	&	\pm1.01	\\
-0.04	&	\pm0.88	&	-2.82	
\end{pmatrix*} $ &
Ti(5,\dots,8) &
$\begin{pmatrix*}[r]					
7.46	&	0.00	&	0.00	\\
0.00	&	5.78	&	0.00	\\
0.00	&	0.00	&	7.29	
\end{pmatrix*} $ 
\\ \\
Ti(1,2) &
$ \begin{pmatrix*}[r]					
6.52	&	\pm0.11	&	0.23	\\
\pm0.37	&	6.60	&	\pm0.06	\\
0.20	&	\pm0.38	&	4.91	
\end{pmatrix*} $ & 
O(9,10) &
$\begin{pmatrix*}[r]
-1.81	&	\pm0.30	&	-0.34	\\
\pm0.29	&	-3.20	&	\pm1.41	\\
-0.35	&	\pm1.31	&	-3.12	
\end{pmatrix*} $ &
O(1,\dots,4) &
$\begin{pmatrix*}[r]					
-5.46	&	0.00	&	0.00	\\
0.00	&	-2.42	&	0.00	\\
0.00	&	0.00	&	-1.68	
\end{pmatrix*} $ 
\\ \\
Ti(3,4) &
$\begin{pmatrix*}[r]					
6.44	&	\pm0.10	&	-0.22	\\
\pm0.30	&	6.39	&	\pm0.01	\\
-0.22	&	\pm0.30	&	4.65	
\end{pmatrix*} $ & 
O(11,12) &
$\begin{pmatrix*}[r]
-1.80	&	\pm0.27	&	0.39	\\
\pm0.27	&	-3.09	&	\pm1.30	\\
0.43	&	\pm1.15	&	-3.11	
\end{pmatrix*} $ &
O(5,\dots,8) &
$\begin{pmatrix*}[r]					
-2.48	&	0.00	&	0.00	\\
0.00	&	-3.86	&	\pm1.36	\\
0.00	&	\pm1.10	&	-3.22	
\end{pmatrix*} $ 
\\ \\
Ti(5,6) &
$\begin{pmatrix*}[r]					
6.35	&	\pm0.96	&	0.14	\\
\pm0.06	&	5.46	&	\pm0.58	\\
-0.15	&	\pm0.21	&	5.95	
\end{pmatrix*} $ & 
O(13,14) &
$\begin{pmatrix*}[r]
-2.29	&	\pm0.24	&	-0.21	\\
\pm0.15	&	-3.86	&	\pm1.44	\\
-0.12	&	\pm1.41	&	-3.25	
\end{pmatrix*} $ &
O(9,\dots,16) &
$\begin{pmatrix*}[r]					
-2.08	&	0.00	&	0.00	\\
0.00	&	-4.16	&	\pm1.91	\\
0.00	&	\pm1.70	&	-3.41	
\end{pmatrix*} $ 
\\ \\
Ti(7,8) &
$\begin{pmatrix*}[r]					
6.15	&	\pm0.88	&	-0.01	\\
\pm0.18	&	5.76	&	\pm0.60	\\
0.25	&	\pm0.10	&	5.30	
\end{pmatrix*} $ & 
O(15,16) &
$\begin{pmatrix*}[r]
-2.38	&	\pm0.28	&	0.19	\\
\pm0.15	&	-3.81	&	\pm1.13	\\
0.13	&	\pm1.24	&	-2.93	
\end{pmatrix*} $ &
O(17,\dots,20) &
$ \begin{pmatrix*}[r]					
-5.99	&	0.00	&	0.00	\\
0.00	&	-1.58	&	0.00	\\
0.00	&	0.00	&	-2.16	
\end{pmatrix*} $ 
\\ \\
 & & O(17,18) &
$\begin{pmatrix*}[r]
-2.04	&	\pm0.19	&	0.16	\\
\pm0.21	&	-3.21	&	\pm0.85	\\
0.23	&	\pm0.84	&	-2.89	
\end{pmatrix*} $ & 
O(21,\dots,28) &
$ \begin{pmatrix*}[r]					
-2.45	&	0.00	&	0.00	\\
0.00	&	-3.24	&	\pm0.78	\\
0.00	&	\pm0.95	&	-3.37	
\end{pmatrix*} $ 
\\ \\
 & & O(19,20) &
$\begin{pmatrix*}[r]
-2.30	&	\pm0.28	&	-0.27	\\
\pm0.34	&	-3.78	&	\pm1.11	\\
-0.38	&	\pm1.11	&	-1.83	
\end{pmatrix*} $ &
& \\ \\
 & & O(21,22) &
$\begin{pmatrix*}[r]
-2.27	&	\pm0.13	&	0.07	\\
\pm0.22	&	-3.06	&	\pm0.71	\\
0.07	&	\pm0.51	&	-3.20	
\end{pmatrix*} $ &
& \\ \\
 & & O(23,24) &
$\begin{pmatrix*}[r]
-2.15	&	\pm0.02	&	-0.07	\\
\pm0.09	&	-2.90	&	\pm0.90	\\
-0.10	&	\pm0.57	&	-3.49	
\end{pmatrix*} $ &
& \\ \\
 & & O(25,26) &
$\begin{pmatrix*}[r]
-4.99	&	\pm0.04	&	-0.10	\\
\pm0.10	&	-1.91	&	\pm0.23	\\
-0.07	&	\pm0.32	&	-1.94	
\end{pmatrix*} $ &
& \\ \\
 & & O(27,28) &
$\begin{pmatrix*}[r]
-5.13	&	\pm0.08	&	0.03	\\
\pm0.04	&	-1.75	&	\pm0.36	\\
0.00	&	\pm0.37	&	-1.99	
\end{pmatrix*} $ &
&  \\
\hline 
\hline
\end{tabular}
}
\end{table*}
\newpage
\begin{table*}
{
\caption{\label{BCval}Pr$_2$Ti$_2$O$_7$ Born effective charge tensors in units of $e$.}
\begin{tabular}{lclclc}
\hline 
\hline
Atom & ${P2_1}$ & Atom & ${P2_1}$ &Atom &  $P2_1/m$ \\
\hline
Pr(1,2) & 
$
 \begin{pmatrix*}[r]
4.01	&	\pm0.03	&	0.02	\\
\pm0.05	&	4.49	&	\pm0.31	\\
-0.05	&	\pm0.21	&	3.17	
 \end{pmatrix*}
$
&
O(1,2) &
$\begin{pmatrix*}[r]					
-5.62	&	\pm0.01	&	0.02	\\
0.00	&	-2.31	&	\pm0.05	\\
0.00	&	\pm0.11	&	-1.29	
\end{pmatrix*} $ &
Pr(1,\dots,4) &
$\begin{pmatrix*}[r]					
4.38	&	0.00	&	0.00	\\
0.00	&	4.22	&	0.00	\\
0.00	&	0.00	&	3.61	
\end{pmatrix*} $ 
\\ \\
Pr(3,4) &
 $ \begin{pmatrix*}[r]					
4.18	&	\pm0.03	&	0.01	\\
\pm0.06	&	4.07	&	\pm0.59	\\
0.03	&	\pm0.04	&	3.75	
\end{pmatrix*} $
 &
 O(3,4) &
$\begin{pmatrix*}[r]
-4.63	&	\pm0.03	&	\pm0.02	\\
0.00	&	-2.45	&	0.09	\\
\pm0.01	&	\pm0.18	&	-1.80	
\end{pmatrix*} $ &
Pr(5,\dots,8) &
$\begin{pmatrix*}[r]	
4.40	&	0.00	&	0.00	\\
0.00	&	4.61	&	0.00	\\
0.00	&	0.00	&	4.18	
\end{pmatrix*}$
\\ \\
Pr(5,6) &
$\begin{pmatrix*}[r]					
4.28	&	\pm0.06	&	-0.10	\\
\pm0.20	&	4.48	&	\pm0.15	\\
-0.08	&	\pm0.64	&	4.39	
\end{pmatrix*}$ &
O(5,6) &
$\begin{pmatrix*}[r]
-2.38	&	\pm0.02	&	-0.06	\\
\pm0.08	&	-3.32	&	\pm1.06	\\
0.05	&	\pm0.97	&	-2.90	
\end{pmatrix*} $ &
Ti(1,\dots,4) & 
$\begin{pmatrix*}[r]					
6.72	&	0.00	&	0.00	\\
0.00	&	8.04	&	0.00	\\
0.00	&	0.00	&	5.44	
\end{pmatrix*}$
\\ \\
Pr(7,8) &
$\begin{pmatrix*}[r]					
4.18	&	\pm0.11	&	0.08	\\
\pm0.16	&	4.60	&	\pm0.28	\\
0.15	&	\pm0.48	&	4.37		
\end{pmatrix*}$ &
O(7,8) &
$\begin{pmatrix*}[r]
-2.41	&	\pm0.01	&	0.04	\\
\pm0.08	&	-3.26	&	\pm0.98	\\
-0.05	&	\pm0.86	&	-2.81	
\end{pmatrix*} $ &
Ti(5,\dots,8) &
$\begin{pmatrix*}[r]					
7.64	&	0.00	&	0.00	\\
0.00	&	5.69	&	0.00	\\
0.00	&	0.00	&	7.44	
\end{pmatrix*} $ 
\\ \\
Ti(1,2) &
$ \begin{pmatrix*}[r]					
6.52	&	\pm0.15	&	0.23	\\
\pm0.41	&	6.54	&	\pm0.10	\\
0.22	&	\pm0.38	&	4.90	
\end{pmatrix*} $ & 
O(9,10) &
$\begin{pmatrix*}[r]
-1.80	&	\pm0.31	&	-0.36	\\
\pm0.29	&	-3.15	&	\pm1.38	\\
-0.36	&	\pm1.31	&	-3.13	
\end{pmatrix*} $ &
O(1,\dots,4) &
$\begin{pmatrix*}[r]					
-5.48	&	0.00	&	0.00	\\
0.00	&	-2.42	&	0.00	\\
0.00	&	0.00	&	-1.67	
\end{pmatrix*} $ 
\\ \\
Ti(3,4) &
$\begin{pmatrix*}[r]					
6.44	&	\pm0.12	&	-0.22	\\
\pm0.34	&	6.33	&	\pm0.04	\\
-0.23	&	\pm0.31	&	4.63	
\end{pmatrix*} $ & 
O(11,12) &
$\begin{pmatrix*}[r]
-1.80	&	\pm0.29	&	0.42	\\
\pm0.28	&	-3.07	&	\pm1.26	\\
0.45	&	\pm1.14	&	-3.11	
\end{pmatrix*} $ &
O(5,\dots,8) &
$\begin{pmatrix*}[r]					
-2.47	&	0.00	&	0.00	\\
0.00	&	-3.88	&	\pm1.36	\\
0.00	&	\pm1.09	&	-3.21	
\end{pmatrix*} $ 
\\ \\
Ti(5,6) &
$\begin{pmatrix*}[r]
6.32	&	\pm1.04	&	0.15	\\
\pm0.07	&	5.42	&	\pm0.59	\\
-0.18	&	\pm0.25	&	5.94	
\end{pmatrix*} $ & 
O(13,14) &
$\begin{pmatrix*}[r]
-2.29	&	\pm0.25	&	-0.22	\\
\pm0.15	&	-3.82	&	\pm1.39	\\
-0.13	&	\pm1.38	&	-3.23	
\end{pmatrix*} $ &
O(9,\dots,16) &
$\begin{pmatrix*}[r]					
-2.08	&	0.00	&	0.00	\\
0.00	&	-4.16	&	\pm1.92	\\
0.00	&	\pm1.70	&	-3.42	
\end{pmatrix*} $ 
\\ \\
Ti(7,8) &
$\begin{pmatrix*}[r]					
6.13	&	\pm0.96	&	-0.01	\\
0.18	&	5.75	&	\pm0.62	\\
0.26	&	\pm0.07	&	5.26	
\end{pmatrix*} $ & 
O(15,16) &
$\begin{pmatrix*}[r]
-2.29	&	\pm0.25	&	-0.22	\\
\pm0.15	&	-3.82	&	\pm1.39	\\
-0.13	&	\pm1.38	&	-3.23	
\end{pmatrix*} $ &
O(17,\dots,20) &
$ \begin{pmatrix*}[r]					
-6.13	&	0.00	&	0.00	\\
0.00	&	-1.53	&	0.00	\\
0.00	&	0.00	&	-2.16	
\end{pmatrix*} $ 
\\ \\
 & & O(17,18) &
$\begin{pmatrix*}[r]
-2.37	&	\pm0.29	&	0.20	\\
\pm0.15	&	-3.78	&	\pm1.09	\\
0.15	&	\pm1.21	&	-2.91	
\end{pmatrix*} $ &
O(21,\dots,28) &
$ \begin{pmatrix*}[r]					
-2.44	&	0.00	&	0.00	\\
0.00	&	-3.21	&	\pm0.77	\\
0.00	&	\pm0.95	&	-3.39	
\end{pmatrix*} $ 
 \\ \\
 & & O(19,20) &
$\begin{pmatrix*}[r]
-2.02	&	\pm0.20	&	0.17	\\
\pm0.21	&	-3.19	&	\pm0.84	\\
0.24	&	\pm0.82	&	-2.88	
\end{pmatrix*} $ &
 \\ \\
 & & O(21,22) &
$\begin{pmatrix*}[r]
-2.25	&	\pm0.15	&	0.06	\\
\pm0.23	&	-3.03	&	\pm0.71	\\
0.08	&	\pm0.48	&	-3.18	
\end{pmatrix*} $ &
& \\ \\
 & & O(23,24) &
$\begin{pmatrix*}[r]
-2.13	&	\pm0.03	&	-0.05	\\
\pm0.09	&	-2.86	&	\pm0.90	\\
-0.10	&	\pm0.55	&	-3.50	
\end{pmatrix*} $ &
& \\ \\
 & & O(25,26) &
$\begin{pmatrix*}[r]
-4.92	&	\pm0.04	&	-0.10	\\
\pm0.10	&	-1.94	&	\pm0.26	\\
-0.07	&	\pm0.35	&	-1.95	
\end{pmatrix*} $ &
& \\ \\
 & & O(27,28) &
$\begin{pmatrix*}[r]
-5.15	&	\pm0.07	&	0.04	\\
\pm0.05	&	-1.75	&	\pm0.38	\\
0.01	&	\pm0.37	&	-1.95		
\end{pmatrix*} $ &
&  \\
\hline 
\hline
\end{tabular}
}
\end{table*}
\newpage
\begin{table*}
{
\caption{\label{BCval}Nd$_2$Ti$_2$O$_7$ Born effective charge tensors in units of $e$.}
\begin{tabular}{lclclc}
Atom & ${P2_1}$ & Atom & ${P2_1}$ &Atom &  $P2_1/m$ \\
\hline
Nd(1,2) & 
$
 \begin{pmatrix*}[r]
3.96	&	\pm0.03	&	0.02	\\
\pm0.05	&	4.47	&	\pm0.31	\\
-0.05	&	\pm0.23	&	3.12	
 \end{pmatrix*}
$
&
O(1,2) &
$\begin{pmatrix*}[r]					
-5.64	&	\pm0.01	&	0.03	\\
0.00	&	-2.30	&	\pm0.07	\\
0.00	&	\pm0.14	&	-1.25	
\end{pmatrix*} $ &
Nd(1,\dots,4) &
$\begin{pmatrix*}[r]					
4.38	&	0.00	&	0.00	\\
0.00	&	4.20	&	0.00	\\
0.00	&	0.00	&	3.59	
\end{pmatrix*} $ 
\\ \\
Nd(3,4) &
 $ \begin{pmatrix*}[r]					
4.14	&	\pm0.03	&	0.01	\\
\pm0.06	&	4.03	&	\pm0.58	\\
0.03	&	\pm0.03	&	3.74	
\end{pmatrix*} $
 &
 O(3,4) &
$\begin{pmatrix*}[r]
-4.56	&	\pm0.03	&	-0.02	\\
\pm0.01	&	-2.45	&	\pm0.11	\\
0.01	&	\pm0.20	&	-1.82	
\end{pmatrix*} $ &
Nd(5,\dots,8) &
$\begin{pmatrix*}[r]	
4.34	&	0.00	&	0.00	\\
0.00	&	4.56	&	0.00	\\
0.00	&	0.00	&	4.13
\end{pmatrix*}$
\\ \\
Nd(5,6) &
$\begin{pmatrix*}[r]					
3.00	&		&		\\
4.25	&	\pm0.06	&	-0.11	\\
\pm0.20	&	4.44	&	\pm0.16	\\
-0.07	&	\pm0.65	&	4.39	
\end{pmatrix*}$ &
O(5,6) &
$\begin{pmatrix*}[r]
-2.36	&	\pm0.01	&	-0.05	\\
\pm0.08	&	-3.28	&	\pm1.03	\\
0.05	&	\pm0.96	&	-2.89	
\end{pmatrix*} $ &
Ti(1,\dots,4) & 
$\begin{pmatrix*}[r]					
6.74	&	0.00	&	0.00	\\
0.00	&	8.08	&	0.00	\\
0.00	&	0.00	&	5.42		
\end{pmatrix*}$
\\ \\
Nd(7,8) &
$\begin{pmatrix*}[r]					
4.11	&	\pm0.11	&	0.09	\\
\pm0.15	&	4.57	&	\pm0.27	\\
0.15	&	\pm0.47	&	4.34	
\end{pmatrix*}$ &
O(7,8) &
$\begin{pmatrix*}[r]
-2.39	&	\pm0.01	&	0.03	\\
0.08	&	-3.22	&	\pm0.96	\\
\pm0.05	&	\pm0.85	&	-2.79	
\end{pmatrix*} $ &
Ti(5,\dots,8) &
$\begin{pmatrix*}[r]					
7.78	&	0.00	&	0.00	\\
0.00	&	5.62	&	0.00	\\
0.00	&	0.00	&	7.53	
\end{pmatrix*} $ 
\\ \\
Ti(1,2) &
$ \begin{pmatrix*}[r]					
6.52	&	\pm0.18	&	0.22	\\
\pm0.44	&	6.48	&	\pm0.13	\\
0.23	&	\pm0.39	&	4.89	
\end{pmatrix*} $ & 
O(9,10) &
$\begin{pmatrix*}[r]
-1.79	&	\pm0.32	&	-0.38	\\
\pm0.29	&	-3.12	&	\pm1.35	\\
-0.37	&	\pm1.31	&	-3.15	
\end{pmatrix*} $ &
O(1,\dots,4) &
$\begin{pmatrix*}[r]					
-5.51	&	0.00	&	0.00	\\
0.00	&	-2.41	&	0.00	\\
0.00	&	0.00	&	-1.66	
\end{pmatrix*} $ 
\\ \\
Ti(3,4) &
$\begin{pmatrix*}[r]					
6.43	&	\pm0.15	&	-0.21	\\
\pm0.38	&	6.28	&	\pm0.06	\\
-0.25	&	\pm0.31	&	4.60	
\end{pmatrix*} $ & 
O(11,12) &
$\begin{pmatrix*}[r]
-1.79	&	\pm0.31	&	0.44	\\
\pm0.28	&	-3.05	&	\pm1.23	\\
0.46	&	\pm1.13	&	-3.11	
\end{pmatrix*} $ &
O(5,\dots,8) &
$\begin{pmatrix*}[r]					
-2.47	&	0.00	&	0.00	\\
0.00	&	-3.89	&	\pm1.36	\\
0.00	&	\pm1.08	&	-3.19	
\end{pmatrix*} $ 
\\ \\
Ti(5,6) &
$\begin{pmatrix*}[r]					
6.31	&	\pm1.10	&	0.14	\\
\pm0.08	&	5.39	&	\pm0.59	\\
-0.20	&	\pm0.28	&	5.93	
\end{pmatrix*} $ & 
O(13,14) &
$\begin{pmatrix*}[r]
-2.29	&	\pm0.26	&	-0.22	\\
\pm0.15	&	-3.78	&	\pm1.36	\\
-0.15	&	\pm1.35	&	-3.21	
\end{pmatrix*} $ &
O(9,\dots,16) &
$\begin{pmatrix*}[r]					
-2.08	&	0.00	&	0.00	\\
0.00	&	-4.16	&	\pm1.93	\\
0.00	&	\pm1.70	&	-3.42	
\end{pmatrix*} $ 
\\ \\
Ti(7,8) &
$\begin{pmatrix*}[r]					
6.11	&	\pm1.03	&	-0.01	\\
\pm0.18	&	5.74	&	\pm0.64	\\
0.27	&	\pm0.05	&	5.23	
\end{pmatrix*} $ & 
O(15,16) &
$\begin{pmatrix*}[r]
-2.36	&	\pm0.29	&	0.20	\\
\pm0.15	&	-3.74	&	\pm1.05	\\
0.17	&	\pm1.18	&	-2.88	
\end{pmatrix*} $ &
O(17,\dots,20) &
$ \begin{pmatrix*}[r]					
-6.24	&	0.00	&	0.00	\\
0.00	&	-1.49	&	0.00	\\
0.00	&	0.00	&	-2.16	
\end{pmatrix*} $ 
\\ \\
 & & O(17,18) &
$\begin{pmatrix*}[r]
-2.00	&	\pm0.20	&	0.17	\\
\pm0.22	&	-3.17	&	\pm0.84	\\
0.24	&	\pm0.81	&	-2.86	
\end{pmatrix*} $ &
O(21,\dots,28) &
$ \begin{pmatrix*}[r]					
-2.42	&	0.00	&	0.00	\\
0.00	&	-3.17	&	\pm0.75	\\
0.00	&	\pm0.94	&	-3.40	
\end{pmatrix*} $ 
 \\ \\
 & & O(19,20) &
$\begin{pmatrix*}[r]
-2.29	&	\pm0.28	&	-0.29	\\
\pm0.35	&	-3.76	&	\pm1.10	\\
-0.42	&	\pm1.10	&	-1.75	
\end{pmatrix*} $ &
 \\ \\
 & & O(21,22) &
$\begin{pmatrix*}[r]
-2.23	&	\pm0.16	&	0.04	\\
\pm0.25	&	-3.00	&	\pm0.70	\\
0.08	&	\pm0.46	&	-3.17	
\end{pmatrix*} $ &
& \\ \\
 & & O(23,24) &
$\begin{pmatrix*}[r]
-2.11	&	\pm0.04	&	-0.03	\\
\pm0.09	&	-2.82	&	\pm0.90	\\
-0.10	&	\pm0.54	&	-3.51	
\end{pmatrix*} $ &
& \\ \\
 & & O(25,26) &
$\begin{pmatrix*}[r]
-4.85	&	\pm0.05	&	-0.10	\\
\pm0.10	&	-1.96	&	\pm0.28	\\
-0.06	&	\pm0.37	&	-1.95		
\end{pmatrix*} $ &
& \\ \\
 & & O(27,28) &
$\begin{pmatrix*}[r]
-5.17	&	\pm0.07	&	0.05	\\
\pm0.06	&	-1.75	&	\pm0.39	\\
0.02	&	\pm0.37	&	-1.90	
\end{pmatrix*} $ &
&  \\
\hline
\hline
\end{tabular}
}
\end{table*}
\end{document}